\documentclass[a4paper,onecolumn,unpublished,amsfonts,amsmath,amssymb,noarxiv]{quantumarticle}
\pdfoutput=1
\usepackage[utf8]{inputenc}
\usepackage[english]{babel}
\usepackage[T1]{fontenc}
\usepackage[unicode=true,
bookmarks=true,bookmarksopen=false,
breaklinks=false,pdfborder={0 0 0},colorlinks=true]
{hyperref}
\usepackage{xcolor}
\definecolor{cblue}{rgb}{0.16, 0.32, 0.75}
\definecolor{cred}{rgb}{0.7, 0.11, 0.11}
\hypersetup{%
	,linkcolor=cred
	,citecolor=cblue
	,urlcolor=cblue
}
\usepackage[numbers,sort&compress]{natbib}
\usepackage{dsfont,enumitem,color,bm,ulem,comment,cancel}
\usepackage[allowbreakbefore,nospacearound,shortcuts]{extdash}

\newcommand{\fbar}[1]{\,\overline{\!#1}}
\newcommand{\fbarsb}[2]{{\,\overline{\!#1}\!}_{#2}}
\newcommand{\sqb}[1]{[\,#1^+\!\!,#1^-]}

\newcommand{\sqm}[1]{\sqb{#1}[\mathcal{D}#1^+][\mathcal{D}#1^-]}
\newcommand{\tup}[2][n]{
    \bm{#2}_{#1}
}

\renewcommand{\part}[1]{
   ~\refstepcounter{part}
    
    \noindent{\Large\sffamily \textbf{Part \Roman{part}}}
    
    \vspace{1em}
    \noindent{\huge\sffamily #1}
    
    \vspace{0.5em}
}
\newcommand{\separator}{
\begin{center}
    \rule{0.75\textwidth}{0.25pt}
\end{center}
}

\numberwithin{equation}{section}
\allowdisplaybreaks		

\usepackage{amsthm}
\theoremstyle{definition}
\newtheorem{theorem}{Theorem}[]
\newtheorem{observation}[theorem]{Phenomenological observation}
					
\begin{document}

\title{Phenomenological quantum mechanics: {\newline I. Phenomenology of quantum observables}}
\author{Piotr Sza{\'n}kowski}
\orcid{0000-0003-4306-8702}
\address{Institute of Physics, Polish Academy of Sciences, al.~Lotnik{\'o}w 32/46, PL 02-668 Warsaw, Poland}
\email{piotr.szankowski@ifpan.edu.pl}

\author{Davide Lonigro}
\orcid{0000-0002-0792-8122}
\address{Department Physik, Friedrich-Alexander-Universität Erlangen-Nürnberg, Staudtstraße 7, 91058 Erlangen, Germany}

\author{Fattah Sakuldee}
\orcid{0000-0001-8756-7904}
\address{Wilczek Quantum Center, School of Physics and Astronomy, Shanghai Jiao Tong University, 800 Dongchuan Road, Minhang, 200240 Shanghai, China}
\address{The International Centre for Theory of Quantum Technologies, University of Gda\'nsk, Jana Ba\.zy\'nskiego 1A, 80-309 Gda\'nsk, Poland}

\author{{\L}ukasz Cywi{\'n}ski}
\orcid{0000-0002-0162-7943}
\address{Institute of Physics, Polish Academy of Sciences, al.~Lotnik{\'o}w 32/46, PL 02-668 Warsaw, Poland}

\author{Dariusz Chru\'{s}ci\'{n}ski}
\orcid{0000-0002-6582-6730}
\address{Institute of Physics, Faculty of Physics, Astronomy and Informatics, Nicolaus Copernicus University,
Grudziadzka 5/7, 87-100 Toru\'n, Poland}

\begin{abstract}
We propose an exercise in which one attempts to deduce the formalism of quantum mechanics solely from phenomenological observations. The only assumed inputs are obtained through sequential probing of quantum systems; no presuppositions about the underlying mathematical structures are permitted. We demonstrate that it is indeed possible to derive, on this basis, a complete and fully functional formalism rooted in the structures of Hilbert spaces. However, the resulting formalism---the \textit{bi}-trajectory formalism---differs significantly from the standard state-focused formulation.

In Part I of the paper, we analyze the outcomes of various experiments involving sequential measurements of quantum observables. These outcomes are quantitatively described by phenomenological multi-time probability distributions, estimated from experimental data. Our first conclusion is that the theory describing these experiments must be non-classical: the measured sequences cannot be interpreted as sampling of a \textit{uni}-trajectory representing the system's observable. The non-classical nature of the investigated systems manifests in a range of observed phenomena, including quantum interference, the quantum Zeno effect, and uncertainty relations between the measured observables.
\end{abstract}


\maketitle

\tableofcontents

\section{Introduction}\label{sec:intro}

One of the iconic features of quantum mechanics is that the standard formalism of the theory implements a dichotomous (two-fold) mode of description. 

The first mode applies in \textit{non-measurement contexts}, where the subject---a quantum system of interest---evolves freely without being disturbed by the probing with \textit{measuring devices} deployed by an observer. In this case, a system is described by its time-dependent \textit{state}, formally represented by a density matrix $\hat\rho(t)$ in the form of a unit-trace positive semi-definite operator acting on a given Hilbert space. The system-specific \textit{dynamical law} governing the state's time evolution is represented in the formalism by a unitary transformation
\begin{align}
    \hat\rho(t_1) = \hat U_{t_1,t_0}\hat\rho(t_0)\hat U_{t_1,t_0}^\dagger.
\end{align}

The second mode is intended for complementary \textit{measurement contexts}, where it is used to describe the perceptions of an observer who intervenes at a given time $t_1$ by deploying a device capable of measuring a system observable. Accordingly, the outcome of such a measurement, as seen by the observer, is formally described by the probability distribution conditioned by the system state, given by the well-known \textit{Born rule},
\begin{align}\label{eq:born}
    P\big(f\mid\hat\rho(t_1)\big) &= \operatorname{tr}[\hat P^F(f)\hat\rho(t_1)].
\end{align}
Here, $\hat F = \sum_{f\in\Omega(F)}f\hat P^F(f)$ is a Hermitian operator representing the measured observable $F$, decomposed in terms of a complete family $\{\hat P^F(f) \mid f\in\Omega(F)\}$ of orthogonal projectors,
\begin{equation}
    \sum_{f\in\Omega(F)}\hat P^F(f) = \hat 1\quad\text{and}\quad\hat P^F(f)\hat P^F(f') = \delta_{f,f'}\hat P^F(f).
\end{equation}

The \textit{collapse rule} supplements the standard formalism by serving as an interface between the two modes. It states that, once the measurement is concluded and one transitions from the measurement context to the non-measurement context, the system is assigned a new state conditioned by the measurement result,
\begin{align}\label{eq:collapse_rule}
    \hat \rho(t_1) \xrightarrow{\text{state collapse}} \hat\rho(t\mid f_1,t_1) 
        = \hat U_{t,t_1}\frac{\hat P^F(f_1)\hat\rho(t_1)\hat P^F(f_1)}{\operatorname{tr}\bigl[\hat P^F(f_1)\hat\rho(t_1)\bigr]}\hat U_{t,t_1}^\dagger,
\end{align}
given that the result at time $t_1$ was $f_1\in\Omega(F)$.

The addition of the collapse rule is necessary to construct the description of a measurement context that extends beyond the first deployment of the measuring device. Hence, in the standard formulation, the observed results of a measurement carried out with devices deployed in a chronological sequence are formally described by chaining the Born rule and the collapse rule: the system state at the time of the subsequent measurement is taken as the state collapsed due to the previous measurement (and evolved for the duration in between the two measurements). That is, the probability that the observer measures the result $f_1$ at time $t_1$, $f_2$ at $t_2$, \ldots, and finally $f_n$ at $t_n$, is given by
\begin{align}\label{eq:born-sequential}
    P_{t_n,\ldots,t_1}(f_n,\ldots,f_1) = P(f_1\mid\hat\rho(t_1))\prod_{j=2}^n P\big(f_j\mid \hat\rho(t_j\mid f_{j-1},t_{j-1};\ldots;f_1,t_1)\big),
\end{align}
where the conditional system states are found by a recursive application of the rule~\eqref{eq:collapse_rule},
\begin{align}
    \hat\rho(t\mid f_j,t_j;\ldots;f_1,t_1) &= \hat U_{t,t_{j}}
        \frac{\hat P^F(f_{j})\hat\rho(t_{j}\mid f_{j-1},t_{j-1};\ldots;f_1,t_1)\hat P^F(f_{j})}{P\big(f_j\mid\hat\rho(t_j\mid f_{j-1},t_{j-1};\ldots;f_1,t_1)\big)}
        \hat U_{t,t_j}^\dagger.
\end{align}

The choice to structure the formalism in this way has certain undesirable consequences. One of them is the necessity for the \textit{Heisenberg cut}~\cite{Atmanspacher_97,Schlosshauer_11,Letertre_21}. The issue is that the two description modes apply only in their respective domains, and so, whenever one wishes to employ the formalism, one has to decide whether a given situation counts as a measurement or not. But what are the quantitative criteria for making such a decision? The current understanding of the capabilities of the standard formalism is that it cannot provide a concrete answer to such a question~\cite{Wigner_95,Frauchiger_18,Kastner_20,Zukowski_21}. Pinpointing the transition between the measurement and non-measurement context for a collection of interacting quantum systems remains an open problem. Therefore, the only available solution is to place the \textit{cut} more or less arbitrarily; as long as one operates far away from the cut, the formalism can be used without further objections. 

However, this is not---and never has been---a satisfactory resolution. On one hand, the friction due to the dichotomous description modes has been an inexhaustible source of inspiration for paradoxical thought experiments (perhaps best exemplified by the Wigner's friend scenarios, in which an observer observes another observer); on the other hand, modern quantum technologies also operate at scales that approach perilously close to the expected location of the Heisenberg cut. Thus, it seems inevitable that the standard formalism, being incapable of describing systems living in the admittedly fuzzy border separating the measurement and non-measurement contexts, will, sooner rather than later, reach the limits of its applicability in cutting-edge quantum physics.

Naturally, over the years, there have been a number of attempts at rectifying these problems. Examples include the de Broglie--Bohm pilot-wave theory~\cite{Bohm_52_1,Bohm_52_2}, Griffiths decoherent histories theory~\cite{Griffiths_JSP84,Gell-Mann_PRD93,Dowker_PRD92,Dowker_JoPA2023}, Everett many-worlds interpretation~\cite{Barrett_12}, and others. However, none of these proposed solutions has garnered universal acceptance as the \textit{de facto} resolution of the tensions identified in the standard formalism. Perhaps the issue lies in the fact that these attempts attack the problem by \textit{postulating} an often sweeping restructuring of the formalism. Even though the proposed modifications are usually motivated by some insightful intuition, in the end they are \textit{ad hoc} propositions.

We concur with the diagnosis that the problems with the standard formalism cannot be reconciled within its framework, and the solution will most likely require a restructuring of the formalism. However, rather than postulating \textit{ad hoc} modifications to the existing formalism, we propose a more systematic approach. The idea is to go back to basics, forget what we know about quantum theory, and attempt to \textit{deduce} the formalism from the experimental observations only---\textit{phenomenological} quantum mechanics. 

We imagine that this route is not dissimilar to the one traversed a century ago by the pioneers of quantum theory, who invented the standard formalism in the first place. However, the breadth of the phenomenology of quantum systems has expanded tremendously over these 100 years; the experimental inputs available to us as a matter of course were considered, back then, as fanciful thought experiments at best and science-fiction at worst. Unsurprisingly, then, the theoreticians of quantum mechanics focused, at its inception, almost exclusively on the phenomenology of single-measurement experiments. This seems to be the main reason why the standard formalism has a dichotomous structure, and the interface between the description mode takes the form of the collapse rule. If this is a correct reading, then the collapse rule appears as an improvised \textit{post hoc} solution that was ``bolted'' onto the formalism because it was just that---a solution to the problem of sequential measurements that, likely, was treated as an afterthought. In that case, the state-centric standard formalism performing sub-optimally in contexts involving sequential measurements or other types of multi-time correlations of quantum observables~\cite{Caves_PRD1986,Aharonov_PRA2009,Sakuldee2018,Milz2020,Milz_PRX2021,taranto_2025,szankowski-2025}, is almost an inevitability. Hence, to avoid the pitfalls encountered by the pioneers, we ought to treat experiments involving sequential measurements as the basic input for our deduction. 

In this two-part series of papers (see~\cite{part2} for Part II), we demonstrate that the phenomenology of sequential measurement experiments is indeed rich enough to deduce a fully functional formalism without presupposing \textit{anything} about its structure (not even a connection with Hilbert spaces has to be assumed). As expected, the so-deduced formalism---the \textit{bi-trajectory formalism}, as we call it---turns out to be structured significantly differently from the standard formalism. In particular, the dichotomous mode of description is absent in the bi-trajectory formulation, allowing it to describe quantum, classical, and in-between systems without invoking the Heisenberg cut, state collapse, or other similar concepts.

Crucially, because the formalism is directly deduced from the phenomenology, its predictions are in full agreement with the empirically confirmed predictions that could also be made with the standard formalism---in this sense, the bi-trajectory formalism is also complete. Additionally, the new formalism can be reduced to a form that is mathematically equivalent to the standard formulation.

The elements of the bi-trajectory formalism have been introduced in previous works~\cite{Szankowski_SciRep20,Szankowski_PRA21,Szankowski_SciPostLecNotes23,Szankowski_Quantum24,Lonigro_Quantum24}. In these works, the bi-trajectory-related concepts have been employed to parameterize open system dynamics~\cite{Szankowski_SciPostLecNotes23}, describe the origins of the classical noise limit in open systems~\cite{Szankowski_SciRep20}, investigate the quantum--to--classical transition~\cite{Szankowski_PRA21,Szankowski_Quantum24}, and in~\cite{Lonigro_Quantum24} all these elements were given solid mathematical foundations. However, only now, with the demonstration of a successful deduction from phenomenology, can the elements of the bi-trajectory formulation be brought together as a holistic formalism for quantum mechanics.

\section{Outline}

In the first part of the paper, we list the phenomenological observations constituting the inputs for the following deduction of the formalism. Section~\ref{sec:sequence_measure} introduces the formal description of the results of experiments involving sequential measurements using multi-time probability distributions and Section~\ref{sec:methodology} explains the methodology for obtaining these probabilities. In Section~\ref{sec:phenomenology}, we use this description to quantify our phenomenological inputs.

We begin in section~\ref{sec:causality} by noting the relation between probability distributions describing sequences of different lengths. The observed relation allows us to conclude that quantum mechanics adheres to a \textit{causal structure} where past affects the future but future cannot affect the past.

Section~\ref{sec:coarse-graining} provides evidence that sequential measurements of a quantum system cannot be interpreted as a sampling of a \textit{trajectory} traced over time by the system's observable. This crucial observation establishes that such systems cannot be described using the classical theory. Since the investigated systems are not constrained by the classical trajectory picture, we recognize that quantum systems might support \textit{inequivalent} observables. We identify an important example in quantum observables measured by devices which probe the same physical quantity but with different resolutions.

Section~\ref{sec:interference} showcases \textit{quantum interference} effects in what amounts to a sequence of $k$-slit generalizations of the classic Young experiment, viewed as an example of sequential measurement with \textit{coarse-grained} devices.

In Section~\ref{sec:independent_sys}, we discuss experiments involving multiple \textit{independent} systems that exhibit quantum interference when measured with devices incapable of making a perfect ``which-subsystem'' distinction. To uphold the spirit of phenomenological quantum mechanics, we introduce, for the purposes of this analysis, the \textit{phenomenological definition} of a composite quantum system.

Section~\ref{sec:markovianity} discusses experimental conditions under which the probability distributions describing sequential measurements exhibit \textit{Markovianity}, enabling us to address the issue of experiment initialization.

Section~\ref{sec:Zeno} describes the quantum Zeno effect, consisting in the following phenomenon: sequences of measuring devices deployed in rapid succession become ``frozen'' in displaying the first measured result when the delay between the deployments tends to zero while their number goes to infinity. This observation is vital for probing the short-time nature of quantum dynamical laws.

Section~\ref{sec:uncertinty_relations} closes the list of phenomenological observations with the \textit{uncertainty relations} between measurements of inequivalent quantum observables.

Section~\ref{sec:bi-traj_picture} shows how the phenomenological observations lead to the conclusion that the emerging formalism is shaping into a \textit{bi-trajectory picture}, a natural progression from the classical ``uni-trajectory picture''.

\section{Description of sequential measurement experiments}\label{sec:sequence_measure}

Assume that the experimenter has access to a collection of \textit{measuring devices} capable of probing observables of a quantum system. When a device measuring an observable $F$---an $F$-device for short---is deployed by the experimenter, one of the possible results $f\in\Omega(F)$,\footnote{
Here, we shall restrict our considerations to $\Omega(F)$ that are finite sets; we defer the case of (countable or uncountable) infinite sets of readouts for future projects.
} corresponding to the measurement outcome obtained by the device at the time $t$ of its deployment, is displayed.

Now, suppose that the experimenter carried out a vast number of sequential measurements with their devices, and they diligently recorded the outcome of each experiment as a sequence (an $n$-tuple) of results,
\begin{align}\label{eq:f_n-tuple}
    \tup{f} = (f_n,\ldots, f_1) \in \Omega(F_n)\times\cdots\times\Omega(F_1) \equiv \Omega(\tup{F}),
\end{align}
displayed by the devices as they were deployed at the corresponding timings,
\begin{align}
    \tup{t} = (t_n,\ldots,t_1)\qquad \text{such that $t_n>\cdots>t_1$.}
\end{align}
These records, obtained for a wide spread of sequence lengths, timings, and device choices, were then used to estimate the corresponding probability distributions,
\begin{align}
    P^{F_n,\ldots,F_1}_{t_n,\ldots,t_1}(f_n,\ldots,f_1) \equiv P^{\tup{F}}_{\tup{t}}(\tup{f}).
\end{align}
Here, $F_j$ at position $j$ in the superscript indicates that the measurement at time $t_j$ was performed with an $F_j$-device. Provided that the experimenter did an outstanding job at finding the most accurate estimations, the family of probabilities $P^{\tup{F}}_{\tup{t}}$ constitutes the complete quantitative description of the phenomenology of sequential measurements.

Consequently, the phenomenological inputs for our deduction of the formalism of quantum mechanics are described using the language of probability theory. Within this language, the set $\Omega(\tup{F})$ from Eq.~\eqref{eq:f_n-tuple} is identified as the \textit{sample space} for an experiment consisting of $n$ measurements where the $F_1$-device was deployed first, followed by the $F_2$-device, \ldots, and terminating with the $F_n$-device. The sequence of results $\tup{f}\in\Omega(\bm F_n)$ recorded as described above is then a \textit{singleton} event, so that the value of the probability distribution for a given $\tup{f}$ equals the \textit{measure} of a set consisting of a single element,
\begin{align}
    P^{\tup{F}}_{\tup{t}}(\tup{f}) = P^{\tup{F}}_{\tup{t}}(\{\tup{f}\})
\end{align}
Note that, with a slight abuse of notation, we are using the same symbol here to denote the probability distribution and the probability measure it is associated with.

Any set of elements of the sample space $\Omega(\tup{F})$, which represents an event, can always be decomposed into a finite union $\cup$ of singletons. For example, take the event $A = \{ \tup{f} \mid f_j \in A_j \subset \Omega(F_j) \}$ representing an \textit{alternative} of singleton events; this set decomposes into the union of singletons as follows,
\begin{align}
    A =  \bigcup_{\tup{f}\in A_n\times\cdots \times A_1} \{\tup{f}\} = \bigcup_{\tup{f}\in A}\{\tup{f}\}.
\end{align}
Given that every probability measure $P^{\tup{F}}_{\tup{t}}$ is by definition additive,
\begin{align}
    P^{\tup{F}}_{\tup{t}}\Big(\bigcup_j A_j\Big) &= \sum_j P^{\tup{F}}_{\tup{t}}(A_j),
    \ \text{for sets such that $A_j\cap A_{j'} =\emptyset$ for all $j\neq j'$},
\end{align}
the singletons $\{ \tup{f}\}$---for which $\{\tup{f}\}\cap\{\tup{f}'\}=\emptyset$ unless $\tup{f} = \tup{f}'$---can therefore be considered as basis events: it suffices to specify the measure of each singleton sequence to completely characterize the probability of any event,
\begin{align}
    P_{\tup{t}}^{\tup{F}}(A) &= P^{\tup{F}}_{\tup{t}}\Big(\bigcup_{\tup{f}\in A}\{\tup{f}\}\Big) =
        \sum_{\tup{f}\in A} P^{\tup{F}}_{\tup{t}}(\{\tup{f}\}) =
        \sum_{\tup{f}\in A} P^{\tup{F}}_{\tup{t}}(\tup{f}).
\end{align}

The probability of measuring any given sequence of results generally depends on the chronology of the performed measurements. Therefore, a result $\tup{f}\in\Omega(\bm F_n)$ obtained in a sequence of $n$ measurements performed at the corresponding times $\tup{t}$ and the same sequence of results but measured at different timings $\tup{t}'\neq \tup{t}$ are formally described as two distinct events. Consequently, the probabilities of those two events are given by two distinct probability measures $P^{\tup{F}}_{\tup{t}}:\operatorname{Pow}(\Omega(\bm F_n))\to[0,1]$ and $P^{\tup{F}}_{\tup{t}'}:\operatorname{Pow}(\Omega(\bm F_n))\to[0,1]$, each corresponding to a unique sequence of timings.\footnote{
    The symbol $\operatorname{Pow}(\Omega(\bm F_n))$ indicates the power set of $\Omega(\bm F_n)$, that is, the set of all subsets of $\Omega(\bm F_n)$. This includes the empty set $\emptyset$, the whole set $\Omega(\bm F_n)$, all singletons $\{\tup{f}\}$, and all their unions.
}

To illustrate these points, consider the following example. From the additivity of probability measures, we obtain the equality
\begin{align}
    \sum_{f_j\in\Omega(F_j)}P^{\tup{F}}_{\tup{t}}(f_n,\ldots,f_j,\ldots,f_1) =
        P^{\tup{F}}_{\tup{t}}\Big(\big\{(f_n,\ldots,f_j,\ldots,f_1)\mid f_j\in\Omega(F_j)\big\} \Big).
\end{align}
In words, this relation means that the probability obtained via marginalization over the result measured at time $t_j$ equals the probability of measuring a sequence $(f_n,\ldots,f_j,\ldots,f_1)$ where all results are set to specific values, except for $f_j$, which can be anything displayed by the $F_j$-device. Intuitively, one might expect that such an alternative, where the sequence is not constrained by any particular value of $f_j$, should be equivalent to an event where the $F_j$-device was never deployed. If this were the case, then the marginalized probability would be equal to probability of a sequence of length $n-1$, in which the entry corresponding to $t_j$ has been struck out,
\begin{align}\label{eq:consistent?}
    \sum_{f_j\in\Omega(F_j)}P^{\tup{F}}_{\tup{t}}(\tup{f})
    \stackrel{?}{=}
    P^{F_n,\ldots,\cancel{F_j},\ldots,F_1}_{t_n,\ldots,\cancel{t_j},\ldots,t_1}
    (f_n,\ldots,\cancel{f_j},\ldots,f_1).
\end{align}
However, since $P^{F_n,\ldots,F_j,\ldots,F_1}_{t_n,\ldots,t_j,\ldots,t_1}$ and $P^{F_n,\ldots,\cancel{F_j},\ldots,F_1}_{t_n,\ldots,\cancel{t_j},\ldots,t_1}$ are measures on distinct event spaces, there is no purely logical or mathematical reason for any relation between probability distributions pertaining to different sequences of timings. If any such relation does exist, it can only be established on the grounds of phenomenological observations of the physical processes they are used to describe. Therefore, whether the aforementioned intuitive postulate (or any other such postulate) is correct must be tested in experiments.

The reason why we said before that the relation of type~\eqref{eq:consistent?} is ``intuitively expected'' is because \textit{consistency conditions}, as they are called, are prominently featured in all \textit{classical} systems~\cite{Accardi1982,Strasberg_PRA19,Smirne_QST19,Milz2020,Sakuldee_PRA21,Strasberg_SciPost2023,Lonigro_Quantum24,Szankowski_Quantum24}:
\begin{align}\label{eq:consistency}
    \sum_{f_j\in\Omega(F_j^\mathrm{cl})}
        P^{\bm F^\mathrm{cl}_n}_{\tup{t}}(\tup{f})
    = P^{\tup{F}^\mathrm{cl}}_{\tup{t}}\big(\{\tup{f}\mid f_j\in\Omega(F_j^\mathrm{cl})\}\big)
    = P^{
        F_n^\mathrm{cl},\ldots,\cancel{F_j^\mathrm{cl}},\ldots,F_1^\mathrm{cl}}_{
        t_n,\ldots,\cancel{t_j},\ldots,t_1
    }(f_n,\ldots,\cancel{f_j},\ldots,f_1),
\end{align}
that holds for all sequences of timings $\tup{t}$, results $\tup{f}$, and observables $\tup{F}^\mathrm{cl}$; the superscript `cl' in $F_j^\mathrm{cl}$ indicates that it is a classical observable measured in a system described with a classical theory. In fact, Eq.~\eqref{eq:consistency} can be considered as the foundational axiom of any classical theory, and the compliance with consistency conditions---as in Eq.~\eqref{eq:consistent?}---has been recognized as a signature of classical behavior in quantum systems \cite{Strasberg_PRA19,Smirne_QST19,Milz2020,Sakuldee_PRA21,Strasberg_SciPost2023,Szankowski_Quantum24}.

Consistency of multi-time probabilities is also a fundamental relation in the mathematical theory of stochastic processes; there, due to the famous Kolmogorov extension theorem~\cite{Kolmogorov,Szankowski_Quantum24,Lonigro_Quantum24}, it implies the existence of the probability measures $\mathcal{P}^{F^\mathrm{cl}}[\,f\,][\mathcal{D}f]$ on the space of \textit{trajectories} $t\mapsto f(t)$. When interpreted in the context of the formal description of classical physical theory (and it is always possible to reformulate every classical theory in this way), trajectories $t \mapsto f(t)$ represent possible histories of a given observable. Then, the distribution $\mathcal{P}^{F^\mathrm{cl}}[\,f\,][\mathcal{D}f]$ can be treated as the \textit{master object} of the theory that fully encapsulates the dynamical properties of the corresponding classical observable $F^\mathrm{cl}$. In particular, the results of sequential measurements of this observable are described by a discrete-time restriction of the trajectory measure (in essence, the inverse of the extension theorem):
\begin{align}
    P^{F^\mathrm{cl}}_{\tup{t}}(\tup{f}) = \int\Big(\prod_{j=1}^n
        \delta_{f_j,f(t_j)}
    \Big)\mathcal{P}^{F^\mathrm{cl}}[\,f\,][\mathcal{D}f].
\end{align}

This classical trajectory picture has profound consequences for the context of measurements. For one, due to assumed consistency of all multi-time probabilities, each classical observable $F^\mathrm{cl}$ is represented by a trajectory $t\mapsto f(t)$ which is continuously defined for every $t$; to put it less formally, all classical observables simultaneously have a definite value. It follows that one can collect all mutually independent observables and treat their trajectories as components of a single vector trajectory---the \textit{elementary} observable $E^\mathrm{cl}$ with probability measure $\mathcal{P}^{E^\mathrm{cl}}[\,e\,][\mathcal{D}e]$. Then, the trajectory representing any observable must be a function of $E^\mathrm{cl}$, and thus, the probability distribution describing its measurements has the form:
\begin{align}
\nonumber
    P^{\tup{F}^\mathrm{cl}}_{\tup{t}}(\tup{f}) &= \int\Bigl\{
        \prod_{j=1}^n p^{F_j^\mathrm{cl}}_{t_j}\big(f_j\mid e(t_j)\big)\Bigr\}
        \mathcal{P}^{E^\mathrm{cl}}[\,e\,][\mathcal{D}e]\\
\nonumber
    &= \sum_{\tup{e}\in\Omega(E^\mathrm{cl})^n}
        \Bigl\{\prod_{j=1}^n p^{F_j^\mathrm{cl}}_{t_j}(f_j\mid e_j)\Bigr\}
        \int\Big(\prod_{k=1}^n\delta_{e_k,e(t_k)}\Big)
        \mathcal{P}^{E^\mathrm{cl}}[\,e\,][\mathcal{D}e]\\
\label{eq:classical_measurement}
    &= \sum_{\tup{e}\in\Omega(E^\mathrm{cl})^n}P^{E^\mathrm{cl}}_{\tup{t}}(\tup{e})
        \prod_{j=1}^n p^{F_j^\mathrm{cl}}_{t_j}(f_j\mid e_j),
\end{align}
where the ``filters'' in the form of conditional probabilities $p_t^{F^\mathrm{cl}}(f|e)$ represent the functional dependence between the trajectories of $F^\mathrm{cl}$ and $E^\mathrm{cl}$. The physical interpretation of Eq.~\eqref{eq:classical_measurement} is that, in classical systems, all observables are \textit{equivalent}: the trajectory representing any observable is a function of the one elementary observable $E^\mathrm{cl}$, and so, the dynamics of $E^\mathrm{cl}$ determines the dynamics of every other observable. Consequently, the measure $\mathcal{P}^{E^\mathrm{cl}}[\,e\,][\mathcal{D}e]$ is the true master object of the classical theory, as it fully characterizes the dynamics of the elementary observable.

In what follows we shall test experimentally whether any of these features of classical theories actually survive in quantum systems, and if they do not, then we will try to determine what replaces them.

\section{Methodology}\label{sec:methodology}

The phenomenological method pursued in this work is based on inputs obtained from experimental observations. Ideally, one would extract these inputs from experiments carried out personally under actual laboratory conditions. However, sourcing the phenomenological inputs from a trustworthy third party---say a competent and reputable professional who has performed their own experiments---should also be regarded as fully acceptable.

Now, the consensus within the physics community is that the agreement between actual experimental observations and the Born rule of standard quantum theory \eqref{eq:born}---including its sequential-measurement variant \eqref{eq:born-sequential}---has been thoroughly tested~\cite{urbasi-2010} and can, for all practical purposes, be considered established. Given this, our position is that a probability calculated according to the Born rule will coincide, to a high degree of accuracy, with the probability estimated from a corresponding series of experimental observations. In this sense, the scientific consensus implies that a Born-rule probability calculated using the standard formalism is as reliable a source of phenomenological input as the results reported by a trustworthy and competent experimentalist.

We therefore believe that this approach remains faithful to the phenomenological ethos---namely, reliance only on what can be observed experimentally---while providing the practical convenience of simulating experiments on paper. The use of theoretically calculated probabilities should therefore be understood as a practical surrogate for the corresponding experimental observations, not as an additional theoretical assumption entering the deduction. For this reason, we do not believe that our methodology undermines the phenomenological character of the program.

A related methodological choice concerns the use of idealized observations. The goal here is not to explain individual experiment results, but rather to identify a minimal collection of phenomenological observations sufficient to fill the gaps in the deduction process. For this reason, we chose to formulate the phenomenological inputs in terms of \textit{idealized} observations, neglecting experimental errors and imperfections. 

At the same time, we do not claim that our choice of observations set is unique or optimal. It is conceivable that an alternative deduction program could be based on a different set of phenomenological observations, perhaps including certain classes of experimental imperfections. In the present work, however, we did not identify any advantage in incorporating such effects at the phenomenological level.

Finally, we would like to emphasize that while the use of the standard Born rule in this manner might suggest that we can only be led to a reconstruction of the standard formalism, the point of the exercise is that it is nevertheless possible to carry out deductive reasoning that ultimately leads to a fundamentally different formulation of the same theory.

\section{Phenomenology of sequential measurement experiments}\label{sec:phenomenology}

\subsection{Causality}\label{sec:causality}
\begin{observation}[Causality]\label{obs:causality}
    When a probability distribution describing the sequential measurement is marginalized over the latest entry in the sequence, one obtains the distribution corresponding to the shorter time sequence:
    \begin{align*}
       \sum_{f_{n}\in\Omega(F_{n})}P^{\tup[n]{F}}_{\tup[n]{t}}(\tup[n]{f}) 
        &= P^{\tup[n-1]{F}}_{\tup[n-1]{t}}(\tup[n-1]{f}).
    \end{align*}
    \separator
\end{observation}

This is the first observation that establishes a relation between phenomenological probability measures over otherwise unrelated event spaces, in particular:
\begin{align}\label{eq:causality}
    P^{\tup[n-1]{F}}_{\tup[n-1]{t}}(\{\tup[n-1]{f}\}) &= 
    \sum_{f_n\in\Omega(F_n)}P^{\tup{F}}_{\tup{t}}(\{\tup{f}\}) 
    =P^{\bm F_n}_{\tup{t}}\Big(\bigcup_{f_n\in\Omega(F_n)}\{\tup{f}\}\Big)
    = P^{\tup[n]{F}}_{\tup{t}}\Big(\big\{\big(\Omega(F_n),\tup[n-1]{f}\big)\big\}\Big),
\end{align}
where we used a shorthand notation for the set
\begin{align}
    \big\{ \big(\Omega(F_n),\tup[n-1]{f}\big)\big\} 
    &:= \{\tup{f}'\mid f'_n\in\Omega(F_n), \text{$f_j' = f_j$ for $j<n$}\}
    = \bigcup_{f_n\in\Omega(F_n)}\{\tup{f}\},
\end{align}
which represents the event ``the sequence of results $\tup[n-1]{f}$ was measured at the corresponding times $\tup[n-1]{t}$, irrespective of the last measurement in the sequence at time $t_n$''.

To recognize the significance of this seemingly trivial property, consider the conditional probability for obtaining the result $f_n$ in the last measurement of the sequence of $n$, conditioned by all the previous measurements; formally, such a probability is given by:
\begin{align}
    P^{F_n|\tup[n-1]{F}}_{t_n|\tup[n-1]{t}}(f_n\mid \tup[n-1]{f}) 
    &= \frac{
        P^{\tup{F}}_{\tup{t}}\Big(
        \big\{\big(f_n,\Omega(\tup[n-1]{F})\big)\big\}\cap
        \big\{\big(\Omega(F_n),\tup[n-1]{f}\big)\big\}
        \Big)
    }{
        P^{\tup{F}}_{\tup{t}}\Big(\big\{\big(\Omega(F_n),\tup[n-1]{f}\big)\big\}\Big)
    }
    = \frac{P^{\tup{F}}_{\tup{t}}(\{\tup{f}\})}{
        P^{\tup{F}}_{\tup{t}}\Big(\big\{\big(\Omega(F_n),\tup[n-1]{f}\big)\big\}\Big)
    },
\end{align}
where $\{(f_n,\Omega(\tup[n-1]{F}))\}$ represents the event ``the result $f_n$ was measured at time $t_n$, irrespective of the previous $n-1$ results in the sequence''. A straightforward computation shows that its intersection with $\{(\Omega(F_n),\tup[n-1]{f})\}$ equals the singleton event,
\begin{align}
\nonumber
    &\big\{\big(f_n,\Omega(\bm F_{n-1})\big)\big\}\cap
    \{(\Omega(F_n),\tup[n-1]{f})\}\\
\nonumber
    &\phantom{=}= \big\{\tup{f}' \mid \text{$\big(f_n' = f_n$ and $f_{j<n}'\in\Omega(F_j)\big)$ and $\big( f_{j<n}' = f_j$ and $f_n'\in\Omega(F_n)\big)$}\big\}\\
    &\phantom{=}= \big\{\tup{f}'\mid \text{$f_n'=f_n$ and $f_{j<n}'=f_j$}\big\}
    = \{ \tup{f} \}.
\end{align}
Now, using the relation~\eqref{eq:causality} we can replace the denominator with the probability distribution describing an experiment consisting of a \textit{shorter} sequence of measurements:
\begin{align}
\label{eq:conditional_prob}
    P^{F_n|\tup[n-1]{F}}_{t_n|\tup[n-1]{t}}(f_n|\tup[n-1]{f}) &=
    \frac{P^{\tup{F}}_{\tup{t}}(\{\tup{f}\})}{
        P^{\tup{F}}_{\tup{t}}\Big(\big\{\big(\Omega(F_n),\tup[n-1]{f}\big)\big\}\Big)
    }
    = \frac{P^{\tup{F}}_{\tup{t}}(\tup{f})}{P^{\tup[n-1]{F}}_{\tup[n-1]{t}}(\tup[n-1]{f})}.
\end{align}
This is a decidedly non-trivial statement as, in an absence of phenomenological property like~\ref{obs:causality}, probabilities conditioned on any other event would not necessarily satisfy analogous relation, e.g.,
\begin{align}\label{eq:non_conditional_prob}
    P^{F_1|F_n,\ldots,F_2}_{t_1|t_n,\ldots,t_2}(f_1|f_n,\ldots,f_2) &=
    \frac{
        P^{\tup{F}}_{\tup{t}}(\{\tup{f}\})
    }{
        P^{\tup{F}}_{\tup{t}}\Big(\big\{\big(f_n,\ldots,f_2,\Omega(F_1)\big)\big\}\Big)
    }
    \stackrel{?}{=} \frac{
        P^{\tup{F}}_{\tup{t}}(\tup{f})
    }{
        P^{F_n,\ldots,F_2}_{t_n,\ldots,t_2}(f_n,\ldots,f_2)
    }.
\end{align}

By implying relation~\eqref{eq:conditional_prob} over other analogous relations like~\eqref{eq:non_conditional_prob}, the observation~\ref{obs:causality} singles out the specific \textit{causal structure} where the past affects the future, but not the vice versa:
\begin{align}
\nonumber
    P^{\tup{F}}_{\tup{t}}(\tup{f}) 
    &= \frac{
            P^{\tup{F}}_{\tup{t}}(\tup{f})
        }{
            P^{\tup[n-1]{F}}_{\tup[n-1]{t}}(\tup[n-1]{f})
        } P^{\tup[n-1]{F}}_{\tup[n-1]{t}}(\tup[n-1]{f})
        =P^{F_n|\tup[n-1]{F}}_{t_n|\tup[n-1]{t}}(f_n\mid\tup[n-1]{f})
        P^{\tup[n-1]{F}}_{\tup[n-1]{t}}(\tup[n-1]{f})\\
\nonumber
    &= P^{F_n|\tup[n-1]{F}}_{t_n|\tup[n-1]{t}}(f_n\mid\tup[n-1]{f})
        P^{F_{n-1}|\tup[n-2]{F}}_{t_{n-1}|\tup[n-2]{t}}(f_2\mid\tup[n-2]{f})
        P^{\tup[n-2]{F}}_{\tup[n-2]{t}}(\tup[n-2]{f})\\
\nonumber
    &\phantom{==}\vdots\\
\label{eq:causal_structure}
    &= P^{F_1}_{t_1}(f_1)\prod_{j=2}^n 
        P^{F_j|\tup[j-1]{F}}_{t_j|\tup[j-1]{t}}(f_j\mid\tup[j-1]{f}).
\end{align}

Physically, the above decomposition indicates that the experimenter is free to choose at any moment to terminate or carry on with their measurements, and their decision, whatever it may be, cannot affect the results they already obtained. In contrast, if causality did not hold, the future could influence the past. To illustrate what this could entail, consider the following scenario. The experimenter is sending us measurement results as they obtain them, but they do not inform us whether they mean to continue the experiment. Without causality, the probability distributions describing the received results would depend on the information the experimenter is withholding from us. This means that, just by examining the results we possess in the present, we would be able to learn whether the experimenter will deploy their devices in the future.

Note that classical theories are also causal in this sense (consistency relations~\eqref{eq:consistency} automatically imply causality). Hence, observation~\ref{obs:causality} is not a special feature of quantum theory. To the contrary, it seems likely that causality is a universal property of physical theories, at least, those that are compatible with the phenomenology the Universe as we know it. However, we are unaware of a proof that non-causal or acausal physical theory could not be constructed; in fact, a class of interpretative frameworks for quantum mechanics invokes \textit{retrocausality}, where the future influences the past in a certain limited sense, as a potential solution to some conceptual issues of the standard formalism \cite{Aharonov_PhysRev1964,Reznik_PRA1995,Aharonov_arxiv2007}.

\subsection{Quantum coarse-grained measurements}\label{sec:coarse-graining}

\begin{observation}[Inconsistency]\label{obs:inconsistency}
When a probability distribution describing the sequential measurement is marginalized over the readouts mid-sequence, it does \textit{not} generally result in a distribution for a shorter sequence:
\begin{align*}
    P^{F_n,\ldots,\cancel{F_j},\ldots, F_1}_{t_n,\ldots,\cancel{t_j},\ldots,t_1}(f_n,\ldots,\cancel{f_j},\ldots,f_1)
        -\sum_{f_j\in\Omega(F_j)}P^{\tup{F}}_{\tup{t}}(\tup{f})
        \neq 0\qquad\text{for }j < n.
\end{align*}
\separator
\end{observation}
The implication is that the multi-time distributions describing measurements of quantum observables \textit{violate} the consistency condition~\eqref{eq:consistency}. Consequently, observation~\ref{obs:inconsistency} confirms that quantum systems must be described with a non-classical theory, because the formal description of measurements is incompatible with the classical trajectory picture. Physically, this means that, in general, a sequence of outcomes obtained by measuring a quantum system \textit{cannot} be interpreted as a sampling of the trajectory traced over time by the system's observable.

If the classical-like trajectory-based formalism is ruled out for quantum mechanics, what kind of formal structures would be compatible with its phenomenology? Notice that when measurements are not constrained by the consistency conditions, like in quantum mechanics, there is no longer a reason for observables to have a definite value at each instant of time. If this is the case, then observables in such a non-classical system must not be equivalent, and thus, not every sequential measurement has to be described by a probability distribution of the form~\eqref{eq:classical_measurement}. This categorical difference between quantum and classical observables can be potentially exploited to uncover the underlying structure of quantum theory.

To this end, consider an experiment in which, at some point in the sequence, instead of deploying the device measuring the observable $F_j$, the experimenter deploys an $\fbarsb{F}{j}$-device that measures the same observable but with less definition, that is, a \textit{coarse-grained} variant of the fine-grained $F_j$-device. In more formal terms, the coarse-graining can be described as follows: given any $\fbar{f}\in\Omega(\fbarsb{F}{j})$, there exists a subset $\omega(\fbar{f})\subseteq\Omega(F_j)$ such that, if the $F_j$-device reads out any of the values $f\in\omega(\fbar f)\subseteq\Omega(F_j)$, then, all else being equal, the $\fbarsb{F}{j}$-device deployed in place of the $F_j$-device displays the result $\fbar{f}\in\Omega(\fbarsb{F}{j})$. The relationship between the coarse- and fine-grained device is then summarized by the \textit{resolution},
\begin{align}\label{eq:resolution_Fbar}
\operatorname{Res}(\fbarsb{F}{j}|F_j) := \Bigl\{\omega(\fbar{f})\subseteq\Omega(F_j)\ \Bigm|\   \fbar f\in\Omega(\fbarsb{F}{j}),\ \forall_{\fbar{f}\neq\fbar{f}'}\omega(\fbar{f})\cap\omega(\fbar{f}')=\emptyset,\ \bigcup_{\fbar f\in\Omega(\fbarsb{F}{j})}\omega(\fbar f) = \Omega(F_j)\Bigr\}.
\end{align}
As the concept of coarse-grained measurements will prove to be a crucial ingredient of various observations, we introduce here a short-hand notation that should make the subsequent analysis more transparent. Let $\omega(\fbarsb{f}{j}) = \{ f_j', f_j'',\ldots,f_j^{(k_j)}\}$ so that $\fbarsb{f}{j}$ is one of the results displayed by $\fbarsb{F}{j}$-device, and $f_j',\ldots,f_j^{(k_j)}$ are the corresponding fine-grained values the device coarse-grains over. Then, for sequences of $\fbarsb{F}{j}$-device readouts, we will use the symbol $\vee$ to explicitly indicate in-line the alternative fine-grained values
\begin{align}
    \Omega(\fbarsb{\bm F}{n})\ni \fbarsb{\bm{f}}{n} &\equiv
    \Big(f_n'\vee\cdots \vee f_n^{(k_n)},\ldots,f_1'\vee \cdots\vee f_1^{(k_1)}\Big)
    = \Big(\bigvee_{\omega(\fbarsb{f}{n})}f_n,\ldots,\bigvee_{\omega(\fbarsb{f}{1})}f_1\Big) = \bigvee_{\omega(\fbarsb{\bm{f}}{n})}\tup{f}.
\end{align}

Physically, the concept of coarse-grained device can be visualized with the following model of measurement. Suppose that the $F$-device operates by first analyzing the measured system into distinct components (e.g., magnetic field gradient splits the beam of spinful particles into spatially dislocated sub-beams), then detecting the presence of individual components (e.g., by scattering the light with short enough wavelength), and finally reversing the analysis step by recombining the components (e.g., inverse the magnetic field gradient). The coarse-grained $\fbar{F}$-device would perform the same pre-detection analysis and post-detection synthesis, but its actual detection method would count a number of components as one individual (e.g., by using light with a longer wavelength).

An experimenter can exploit the classical-like causality property (observation~\ref{obs:causality}) to test whether the given $\fbar{F}$-device does indeed operate as a coarse-grained version of the $F$-device---i.e., whether the resolution~\eqref{eq:resolution_Fbar} can be attributed to the device---by verifying if the corresponding probability distributions satisfy the condition
\begin{align}\label{eq:coarse-grained_measurement}
    \sum_{f\in\omega(\fbar{f})}P^F_t(f) = P_t^{\fbar{F}}\left(\fbar{f}\,\right)\quad\text{for every $\fbar{f}\in\Omega(\fbar{F})$ and $t>0$}.
\end{align}

However, this test only checks for a necessary, and not sufficient, condition. To understand this point, let us consider how coarse-grained measurements are achieved for classical observables. If we denote the classical coarse-grained observable as $\widetilde{F}$, then, since the trajectory representing $\widetilde{F}$ is a function of the trajectory of $F$ (which, in turn, is a function of the elementary trajectory), the effect of coarse-graining is obtained by setting the filter functions in Eq.~\eqref{eq:classical_measurement} to $p^{\widetilde{F}}_t(\fbar{f}|f) = \sum_{f'\in\omega(\fbar{f})}\delta_{f,f'}$, so that
\begin{align}\label{eq:faux-coarse-grained_measurement}
    P^{\widetilde{F}}_{\tup{t}}(\fbarsb{\bm{f}}{n}) &
        = \sum_{\tup{f}}P_{\tup{t}}^F (\tup{f})\prod_{j=1}^n p^{\widetilde{F}}_{t_j}(\fbarsb{f}{j}|f_j)
        = \sum_{\tup{f}}P_{\tup{t}}^F (\tup{f})\prod_{j=1}^n \sum_{f_j'\in\omega(\fbarsb{f}{j})}\delta_{f_j,f_j'}
        = \sum_{\tup{f} \in \omega(\fbarsb{\bm{f}}{n})} P^F_{\tup{t}}(\tup{f}).
\end{align}
Not only such a device satisfies the necessary condition~\eqref{eq:coarse-grained_measurement}, it is also the only classical device that does. Consequently, when one performs a sequential measurement of a classical observable, the readout of the coarse-grained $\widetilde{F}$-device is equivalent to the \textit{alternative} of readouts of the fine-grained variant of the device,
\begin{align}\label{eq:faux-coarse-grained_k-slit}
    P^{F,\ldots,\widetilde{F},\ldots,F}_{t_n,\ldots,t_j,\ldots,t_1}\Big(f_n,\ldots,\bigvee_{\omega(\fbarsb{f}{j})}f_j,\ldots,f_1\Big)
        = \sum_{f_j\in\omega(\fbarsb{f}{j})}P^{F,\ldots,F,\ldots,F}_{t_n,\ldots,t_j,\ldots,t_1}(f_n,\ldots,f_j,\ldots,f_1).
\end{align}
Such a \textit{faux}-coarse-grained device can also be implemented for quantum observables: the quantum version of the $\widetilde{F}$-device works by first performing the actual measurement with the fine-grained $F$-device, and then displaying the post-processed results to the experimenter; formally, this amounts to using probability distribution of the form~\eqref{eq:classical_measurement}, in which the filter function describes the post-processing. With the post-processing set to $p^{\widetilde{F}}_t$, the result of a sequential measurement using $\widetilde{F}$-device would be described by Eq.~\eqref{eq:faux-coarse-grained_measurement}.

However, since quantum observables are not constrained by the classical trajectory picture, it is physically possible (or rather, it is not forbidden as a matter of principle) to construct a \textit{quantum} coarse-graining $\fbar{F}$-device which, on one hand, satisfies condition~\eqref{eq:coarse-grained_measurement}, and on the other hand, \textit{cannot} be described by a distribution of the form~\eqref{eq:faux-coarse-grained_measurement}. To phrase it less formally: A proper quantum coarse-grained $\fbar{F}$-device must perform its measurement via physical processes that are \textit{distinct} from those utilized by the $F$-device. Otherwise, the action of the $\fbar{F}$-device would be physically equivalent to the action of the $F$-device, modulo some form of post-processing of the obtained data, which would render it a faux-coarse-grained device instead.

\begin{observation}[Quantum coarse-grained measurement]\label{obs:coarse-grained_measurement} 
When a proper quantum coarse-graining $\fbarsb{F}{j}$-device, characterized by a resolution $\operatorname{Res}(\fbarsb{F}{j}|F_j)$, is deployed mid sequence of measurements at time $t_j < t_n$, we observe that
\begin{align}\label{eq:resolution_crit}
    P_{t_n,\ldots,t_j,\ldots,t_1}^{F_n,\ldots,\fbarsb{F}{j},\ldots,F_1}\Big(
        f_n,\ldots,\bigvee_{\omega(\fbarsb{f}{j})}f_j,\ldots,f_1\Big) 
    - \sum_{f_j\in\omega(\fbarsb{f}{j})}P_{\tup{t}}^{\tup{F}}(\tup{f}) \neq 0,
\end{align}
that is, the readout of a coarse-grained device is \textit{not} equivalent to the alternative of fine-grained readouts. However, the classical picture always persists for the latest readout of the sequence due to causality (observation~\ref{obs:causality}),
\begin{align*}
    P^{\fbarsb{F}{n},F_{n-1},\ldots,F_1}_{t_n,t_{n-1},\ldots,t_1}\Big(\bigvee_{\omega(\fbarsb{f}{n})}f_n,f_{n-1},\ldots,f_1\Big)
        = \sum_{f_n\in\omega(\fbarsb{f}{n})}P^{\tup{F}}_{\tup{t}}(\tup{f}).
\end{align*}
\separator
\end{observation}
This observation simultaneously constitutes an operational criterion for distinguishing a faux-coarse-grained device from genuine quantum coarse-grained device: a device satisfying the necessary condition \eqref{eq:coarse-grained_measurement} is not a faux-coarse-grained device---and thus, is quantum coarse-grained---when it also satisfies \eqref{eq:resolution_crit}.

We finally remark that, at least in some cases, it is possible to manipulate the physical construction of a quantum coarse-grained device to effectively adjust its resolution. In the example discussed previously, such a manipulation would amount to varying the wavelength of the light used for detection---the longer the wavelength, the more coarse-grained the device. This idea can be pushed to its logical extreme, where we observe the quantum analogue of the classical consistency relation:
\begin{observation}[Extreme quantum coarse-graining]\label{obs:extrm_coarse} Let the $\fbarsb{F}{j}$-device be an extremely coarse-grained variant of the $F_j$-device, so that it is incapable of resolving between any alternatives: $\operatorname{Res}(\fbarsb{F}{j}|F_j) =\{\Omega(F_j)\}$. Then, deploying this device at any time in the sequence yields the same results as if no device was deployed at that time,
\begin{align*}
    P_{t_n,\ldots,t_j,\ldots,t_1}^{F_n,\ldots,\fbarsb{F}{j},\ldots,F_1}\Big(
        f_n,\ldots,\bigvee_{\Omega(F_j)}f_j,\ldots,f_1\Big)
    &=P_{t_n,\ldots,\cancel{t_j},\ldots,t_1}^{F_n,\ldots,\cancel{F_j},\ldots,F_1}(
        f_n,\ldots,\cancel{f_j},\ldots,f_1).
\end{align*}
\separator
\end{observation}

Given that quantum coarse-grained devices can be built with adjustable resolutions, we will generally use the symbol $\fbar{F}$ to represent a \textit{class} of measuring devices rather than a specific device. Members of this class are all coarse-grained variants of the $F$-device, with each device characterized by a particular resolution. From this point onward, unless otherwise stated, when we write $P^{\,\ldots \fbarsb{F}{j} \ldots}_{\ldots t_j \ldots}(\ldots \bigvee_{\omega(\fbarsb{f}{j})} f_j \ldots)$, we mean that at time $t_j$, the experimenter deployed one of the devices from the coarse-grained class, with the resolution partially defined by the argument $\bigvee_{\omega(\fbarsb{f}{j})}f_j$. Although this notation does not fully specify the resolution of the deployed device—since it provides only one element, $\omega(\fbarsb{f}{j}) \in \operatorname{Res}(\fbarsb{F}{j}|F_j)$—in most cases, this will be sufficient for our purposes. If circumstances require an unambiguous determination of the resolution, it will be explicitly stated.

\subsection{Quantum interference}\label{sec:interference}

The effect revealed by the coarse-grained measurements described in observation~\ref{obs:coarse-grained_measurement}, consisting in quantum observable not conforming to the classical picture of a trajectory tracing through alternative routes, is an example of the emblematic phenomenon of \textit{quantum interference}. Indeed, an experiment in which a coarse-grained $\fbar{F}$-device is deployed in place of the fine-grained $F$-device can be seen as an instance of a $k$-slit Young experiment analyzed as sequential measurements.

Traditionally, a $k$-slit experiment is presented as involving only a single measurement at its terminus~\cite{Feynman_05}. The slits are then viewed as ``alternative routes'' the measured system can take to reach the measuring device that concludes the experiment; the quantum interference between those alternatives is pointed to as the cause of the measured interference pattern. The $k$-slit experiment is contrasted with a hypothetical scenario in which the experimenter knows---or could even potentially know---which slit the system has passed through on its route to the measuring device. It is then asserted that possessing such a knowledge (or even the potential to acquire it) would prevent the interference from occurring, allowing the results to be described as a sum of the probabilities of each alternative.

In the treatment presented here, this textbook summary of interference experiments can be reformulated with a more precise meaning. First, the notion of the experimenter ``possessing the knowledge of which slit was taken'' corresponds to performing a fine-grained measurement with the $F_j$-device, with the result $f_j\in\omega(\fbarsb{f}{j})\subset\Omega(F_j)$ indicating the slit taken. The ``potential to know'', on the other hand, corresponds to the measurement with a faux-coarse-grained $\widetilde{F}_j$-device, which was discussed in Section~\ref{sec:coarse-graining}. In both scenarios there is, indeed, no interference effect. In the latter case, this is obvious from Eq.~\eqref{eq:faux-coarse-grained_k-slit}. In the former case, the event corresponding to the experimenter performing measurements while knowing that $f_j$ was one of the values in $\omega(\fbarsb{f}{j})$ is represented by the set $\{(f_n,\ldots,f_j,\ldots,f_1)\mid f_j\in\omega(\fbarsb{f}{j})\} = \bigcup_{f_j\in\omega(\fbarsb{f}{j})}\{\tup{f}\}$, and thus
\begin{align}
    P^{\tup{F}}_{\tup{t}}\Big(\bigcup_{f_j\in\omega(\fbarsb{f}{j})}\{\tup{f}\}\Big) = \sum_{f_j\in\omega(\fbarsb{f}{j})}P^{\tup{F}}_{\tup{t}}(\tup{f}).
\end{align}
Finally, since the quantum coarse-grained $\fbarsb{F}{j}$-device, for which we observe interference, is neither an $F_j$-device nor a $\widetilde{F}_j$-device, deploying this device cannot count as ``acquiring knowledge'' or ``having the potential to know'' which slit the system passed through. This formulation also underscores the non-classical nature of the effect, as argued in Section~\ref{sec:coarse-graining}, where it is demonstrated that the proper coarse-grained $\fbarsb{F}{j}$-device is a quintessentially quantum concept with no direct classical analogue.

Further experimentation with quantum coarse-grained devices reveals a crucial feature of quantum interference:
\begin{observation}[Pair-wise quantum interference]\label{obs:pair-wise_interference} The results of a $k$-slit experiment are fully explained with the results of double-slit experiments, indicating that the quantum interference occurs only between \textit{pairs} of alternatives. In particular, it is found that the results of experiments involving sequential measurements with quantum coarse-grained devices can be described using the results obtained with devices that coarse-grain, at most, pairs of fine-grained outcomes. 

It is found that the explicit decomposition of an arbitrary probability distribution into a combination of distributions involving, at most, pair-wise coarse-graining is obtained by applying the following recurrence rule:
\begin{align*}
        &P^{\,\ldots \fbarsb{F}{j} \ldots}_{\ldots t_j \ldots}\Big(\ldots\bigvee_{\omega(\fbarsb{f}{j})}f_j\ldots\Big) 
            = \sum_{f_j\in\omega(\fbarsb{f}{j})}P^{\,\ldots F_j \ldots}_{\ldots t_j\ldots}(\ldots f_j \ldots)\\
            &\phantom{==}+\sum_{\substack{f^\pm_j\in\omega(\fbarsb{f}{j})\\f_j^+\neq f_j^-}}\frac{1}{2}\Big[
                P^{\,\ldots\fbarsb{F}{j}\ldots}_{\ldots t_j \ldots}(\ldots f^+_j\vee f_j^- \ldots )
                - P^{\,\ldots F_j \ldots}_{\ldots t_j \ldots}(\ldots f_j^+ \ldots )
                - P^{\,\ldots F_j \ldots}_{\ldots t_j \ldots}(\ldots f_j^- \ldots )
            \Big],
    \end{align*}
which is applicable for $j<n$; the rule for $j=n$ case is described in observation~\ref{obs:coarse-grained_measurement}.
\separator
\end{observation}

The decomposition obtained through the recurrence described in observation~\ref{obs:pair-wise_interference} becomes unwieldy and opaque when considering sequences with more than a couple of coarse-grained measurements. Therefore, instead of insisting on expressing everything only in terms of the phenomenological probability distributions, it is convenient to parameterize those probabilities with a formal element that allows one to display the structure of the decomposition with more transparency. To this end, we shall posit that the probability distributions can be written as
\begin{align}\label{eq:diagonal_interference_term}
    P^{\tup{F}}_{\tup{t}}(\tup{f}) \equiv I^{\tup{F}}_{\tup{t}}(\tup{f} , \tup{f} ),
\end{align}
where the functions $I^{\tup{F}}_{\tup{t}}:\Omega(\bm F_n)\times\Omega(\bm F_n)\to \mathbb{R}$ are defined to be symmetric with respect to exchange of their arguments $\tup{f}^+\leftrightarrow\tup{f}^-$,
\begin{align}
    I^{\tup{F}}_{\tup{t}}(\tup{f}^+,\tup{f}^-) = I^{\tup{F}}_{\tup{t}}(\tup{f}^- ,\tup{f}^+).
\end{align}
We assume nothing about observables in~\eqref{eq:diagonal_interference_term}, therefore the parameterization works for both fine- and coarse-grained devices. However, to successfully recreate the recurrence rule from observation~\ref{obs:pair-wise_interference}, the function $I^{\tup{F}}_{\tup{t}}$ must also exhibit additivity with respect to the coarse-graining $\vee$ treated as a formal operation on sequences, so that
\begin{align}\label{eq:quantum_interference}
    P^{\fbarsb{\bm{F}}{n}}_{\tup{t}}(\fbarsb{\bm{f}}{n})
    = P^{\fbarsb{\bm{F}}{n}}_{\tup{t}}\Big(\bigvee_{\omega(\fbarsb{\bm{f}}{n})}\tup{f}\Big)
    = I^{\fbarsb{\bm{F}}{n}}_{\tup{t}}\Big(
        \bigvee_{\omega(\fbarsb{\bm{f}}{n})}\tup{f}^+,\bigvee_{\omega(\fbarsb{\bm{f}}{n})}\tup{f}^-
    \Big)
    \equiv \sum_{\tup{f}^\pm \in \omega(\fbarsb{\bm{f}}{n})} I^{\tup[n]{F}}_{\tup{t}}(\tup{f}^+,\tup{f}^-).
\end{align}

To demonstrate its veracity, we now apply this parameterization and recover the recurrence rule:
\begin{align}
\nonumber
    &\sum_{f_j\in\omega(\fbarsb{f}{j})}P^{\,\ldots F_j \ldots}_{\ldots t_j\ldots}(\ldots f_j \ldots)\\
\nonumber
     &\phantom{=}+\sum_{\substack{f_j^\pm\in\omega(\fbarsb{f}{j})\\f_j^+\neq f_j^-}}
         \frac{1}{2}\Big[
             P^{\,\ldots\fbarsb{F}{j}\ldots}_{\ldots t_j\ldots}(\ldots f^+_j\vee f_j^-\ldots)
             - P^{\,\ldots F_j \ldots}_{\ldots t_j \ldots }(\ldots f_j^+ \ldots) - P^{\,\ldots F_j \ldots}_{\ldots t_j \ldots}(\ldots f_j^-\ldots)\Big]\\
\nonumber
    &=\sum_{f_j\in\omega(\fbarsb{f}{j})}I^{\,\ldots F_j \ldots}_{\ldots t_j \ldots}(
        \ldots f_j \ldots\,;\,\ldots f_j \ldots
    )\\
\nonumber
    &\phantom{=}+ \sum_{\substack{f_j^\pm\in\omega(\fbarsb{f}{j})\\f^+_j\neq f_j^-}}\frac{1}{2}\Big[
        I^{\,\ldots F_j\ldots}_{\ldots t_j \ldots}(\ldots f_j^+ \ldots\,;\,\ldots f_j^-\ldots )
        + I^{\,\ldots F_j\ldots}_{\ldots t_j \ldots}(\ldots f_j^- \ldots\,;\,\ldots f_j^+\ldots )
        \Big]\\
\nonumber
    &=
        \sum_{f_j^\pm\in\omega(\fbarsb{f}{j})}
        I^{\,\ldots F_j\ldots}_{\ldots t_j \ldots}(\ldots f_j^+ \ldots\,;\,\ldots f_j^-\ldots )
            = I_{\ldots t_j \ldots}^{\,\ldots \fbarsb{F}{j} \ldots}\Big(\ldots\bigvee_{\omega(\fbarsb{f}{j})}f_j^+\ldots\,
                ;\,\ldots \bigvee_{\omega(\fbarsb{f}{j})}f_j^- \ldots\Big)\\
    &\phantom{=}= P^{\,\ldots\fbarsb{F}{j}\ldots}_{\ldots t_j \ldots}\Big(\ldots\bigvee_{\omega(\fbarsb{f}{j})}f_j \ldots\Big).
\end{align}

The added value in using the functions $I_{\tup{t}}^{\tup{F}}$ is the possibility to reformulate the content of observation~\ref{obs:pair-wise_interference} as an explicit decomposition that reduces all instances of the overt coarse-grained measurements in terms of relatively fine-grained measurements. It has been pointed out~\cite{Sorkin_ModPhysLettA94,Sorkin_JoPA07} that the pair-wise interference between routing alternatives could be considered as a defining feature of the quantum theory. The parameterization~\eqref{eq:quantum_interference} reflects this fact via the structure of the \textit{interference terms} $I_{\tup{t}}^{\tup{F}}$ as functions of \textit{pairs} of sequences, specifically. In principle, one could imagine a non-classical theory (i.e., a physical theory not describable within the trajectory picture) in which the interference also occurs between $k>2$ alternatives. As we shall demonstrate later down the line, the fact that $k=2$ in quantum mechanics has profound consequences for the whole structure of its formalism. Undoubtedly, the same would be true for hypothetical theories with $k>2$, which would likely be even more ``exotic'' than quantum mechanics itself, provided they could exist as physical theories.

\subsection{Quantum interference between independent systems}\label{sec:independent_sys}

The analysis of quantum interference suggests that it is possible to parameterize the formal description of experimental results with (non-positive valued) distributions on the space of pairs of sequences of outcomes, associated with a given quantum system---or, at the very least, with observables of that system.
Indeed, the family of interference terms $I^{\tup{F}}_{\tup{t}}:\Omega(\bm F_n)\times\Omega(\bm F_n)\to\mathbb{R}$ are such distributions, and they seem to be capable to formally describe the sequential measurement outcomes, including the interference effects manifesting in (sequential) $k$-slit experiments. This is certainly the case as long as \textit{one} system is probed by the experimenter. We will now consider certain type of quantum interference in experiments where \textit{multiple} systems are measured at the same time---as it turns out, the description with interference terms alone will be incomplete.

Although the concept of a system composed of a number of subsystems is an intuitive one, we should start by establishing a \textit{phenomenological} definition, that is, one that refers only to measurements: A system will be said to be \textit{partitionable} (into subsystems) when it is possible to deploy \textit{at the same time} the distinct measuring devices that could also be deployed individually to measure the corresponding observables one at the time. If this is the case, then each of those observables can be attributed to one of the subsystems the system is partitionable into.

Some clarifying comments are in order, in case one cannot fully immerse themselves in the role of someone completely ignorant of the standard formalism of quantum mechanics and take the above definition at the face value. The following points should be kept in mind when trying to compare the concept of partitionable system with the standard formal definition of composite system: 
\begin{enumerate}[label=\textnormal{(\roman*)}]
\item When deploying measuring devices, one has to take into account things like the physical size of the device and its structural integrity. In particular, the devices mentioned in the definition above must be built as separate constructs---otherwise they could not be deployed individually. 

\item The notion of ``deployment of measuring devices \textit{at the same time}'' might seem to echo the notion of ``\textit{simultaneous} measurements of commuting observables'' from the standard formulation; however, this is not the case. In accordance with the measurement--projector link in the standard formalism (we shall deduce such a link later in the second part of the paper), a measuring device corresponds to the partitioning of the Hilbert space of the system into the images of a complete set of orthonormal projectors. Therefore, when the device is deployed, it measures any observable whose eigenspaces (defined by the operator representing it in the standard formalism) overlap with the space partitioning of the device. Of course, ``commuting observables'' have the same eigenspaces, and thus, in this sense, the deployment of a \textit{single} device measures all of them \textit{simultaneously}. However, performing the simultaneous measurement in the sense of the phenomenological definition means that \textit{two} (or more) devices have to be deployed \textit{at the same time}. In the case of commuting observables---i.e., observables measured by one kind of device---, this would mean deploying a number of identical copies of one device, which is impossible because they would have to occupy the same volume. 

\item A way to circumvent the issue of devices occupying the same physical space would be to augment them with a mechanism to spatially split and divert the system to make room for their deployment. However, the fact that one is able to split a system into parts that can then be simultaneously measured by a number of devices only enforces the idea that the system is partitionable. 

\item Coarse- and fine-grained observables are a special case of commuting observables: a device performing the simultaneous fine- and coarse-grained measurement is actually equivalent to fine-grained measurement, and thus, it counts as a faux-coarse-grained device. 

\item Finally, it is possible that the proposed definition does not exactly match the notion of composite system from the standard formalism where it is defined as a system represented by a product of Hilbert spaces (whence the choice to use the name ``partitionable'' rather than ``composite''). However, even if it were true, it would not be problematic, because this phenomenological definition leads to the definition of independent systems that does overlap with the notion from the standard formalism.
\end{enumerate}

Now, suppose that the experimenter has access to a system partitionable into two subsystems, $A$ and $B$. For brevity, we shall use the same symbols to denote the two observables that determine the system partitioning (as per the definition given above); and thus, we have an $A$- and a $B$-device that can be deployed simultaneously to measure observables belonging to the corresponding subsystems. The results measured by those devices are then described by the sample spaces $\{\tup{a}\mid a_j\in\Omega(A)\}$ and $\{ \tup{b} \mid b_j\in\Omega(B)\}$. 
 
\begin{observation}[Simultaneous measurements of independent systems]\label{obs:independent_subsystems} Certain systems are found to be partitionable into \textit{independent} subsystems. A trivial example would be a system for which the measurements defining the partitioning are performed in two remote laboratories. 

The constituent subsystems of a partitionable system shall be considered as \textit{independent} when the results of the measurements of any subsystem observable are indistinguishable from measurements of the same observable performed when the measured subsystem is the only present system. In terms of probability distributions, $A$ and $B$ are independent when
\begin{align*}
    P^{AB}_{\tup{t}}(a_n\wedge b_n,\ldots,a_1\wedge b_1) &= P^{AB}_{\tup{t}}(\tup{a}\wedge\tup{b}) = P^{A}_{\tup{t}}(\tup{a})P^B_{\tup{t}}(\tup{b}),
\end{align*}
where $a_j \wedge b_j$ indicate that the $A$-device and $B$-device were deployed simultaneously---an $AB$-device for short---at time $t_j$, and $P^{A/B}_{\tup{t}}$ is the probability distribution describing the measurement with the $A/B$-device in an experiment where only one system (and thus one observable) is present. 
\separator
\end{observation}

As we can see, a distribution associated with a system composed of independent subsystems \textit{factorizes} into distributions associated with individual subsystems. Nevertheless, the subsystem-only interference terms turn out to be insufficient to parameterize all probability distributions describing experiments involving quantum interference in systems partitionable into independent parts. To observe this, one has to deploy a coarse-grained $\fbar{AB}$-device, a version of the device that measures $A$ and $B$ simultaneously but is incapable of making a perfect ``which subsystem'' distinction. Such a device must not be confused with the $\fbar{A}\fbar{B}$-device, consisting in the deployment of the coarse-grained $A$- and $B$-only devices in tandem while the distinction between subsystems is always precise.

\begin{observation}[Quantum interference between independent systems]\label{obs:which-subsys_interference}
Given a quantum coarse-grained $\fbar{AB}$-device with resolution
\begin{equation*}
    \operatorname{Res}(\fbar{AB}|AB) = \Bigl\{
        \omega(\overline{a{\wedge}b}) \subseteq \Omega(A)\times\Omega(B)\Bigm| 
        \bigcup_{\overline{a{\wedge}b}\in\Omega(\overline{AB})}\omega(\overline{a{\wedge}b})
            =\Omega(A)\times\Omega(B)
    \Bigr\},
\end{equation*}
one finds that quantum observables $A$ and $B$ of independent subsystems interfere with each other when the $\fbar{AB}$-device, which is incapable of a perfect distinction between subsystems, is deployed in place of an $AB$-device that resolves subsystems perfectly,
\begin{align*}
    &I^{\,\ldots AB \ldots}_{\ldots t_j \ldots}(
        \ldots a_j^+\!\wedge b_j^+\ldots\,;\,\ldots a_j^-\!\wedge b_j^- \ldots )\\
    &\phantom{=}=
        \frac{1}{2}\Big[P^{\,\ldots \fbar{AB} \ldots}_{\ldots t_j \ldots}\Big(
        \ldots (a^+_j\!\wedge b^+_j)\vee(a_j^-\!\wedge b_j^-) \ldots \Big)\\
    &\phantom{==\frac{1}{2}\Big[}
        - P^{\,\ldots AB \ldots}_{\ldots t_j\ldots}(\ldots a_j^+\!\wedge b_j^+ \ldots)
        - P^{\,\ldots AB \ldots}_{\ldots t_j\ldots}(\ldots a_j^-\!\wedge b_j^- \ldots)\Big]\\
    &\phantom{=}\neq 0.
\end{align*}
However, even though the subsystems $A$ and $B$ are independent, this interference cannot be fully explained exclusively in terms of subsystem interference terms. In formal terms, the \textit{co-interference term}
\begin{align*}
    \Phi^{\,\ldots AB \ldots}_{\ldots t_j \ldots}(
        \ldots a_j^+\!\wedge b_j^+\ldots\,;\,\ldots a_j^-\!\wedge b_j^-\ldots
    )&:= I^{\,\ldots AB \ldots}_{\ldots t_j \ldots}(
        \ldots a_j^+\!\wedge b_j^+\ldots\,;\,\ldots a_j^-\!\wedge b_j^-\ldots)\\
    &\phantom{==}-
        I^{\,\ldots A \ldots}_{\ldots t_j \ldots}(\ldots a^+_j \ldots\,;\,\ldots a^-_j \ldots)\\
    &\phantom{===}
        \times I^{\,\ldots B \ldots}_{\ldots t_j \ldots}(\ldots b^+_j \ldots\,;\,\ldots b^-_j \ldots),
\end{align*}
is found to be non-zero in general.
\separator
\end{observation}

The following observation confirms that the factorization rule has not been violated here. It also provides clues about the form of the actual distribution that can correctly parameterize the ``which subsystem'' interference experiments. 
\begin{observation}[Interference between identical independent systems]\label{obs:identical_subsystems}
Consider the ``which subsystem'' interference experiment carried out on a system partitionable into two independent and otherwise identical subsystems. Let the measurement causing the interference effect be performed with an $\fbar{FF}$-device, which is a coarse-grained variant of the $FF$-device which simultaneously measures the same observable $F$ in both subsystems (in particular, $F=A$ or $F=B$). 

Then, when the co-interference terms $\Phi^{\,\ldots AA \ldots}_{\ldots t_j \ldots}$ and $\Phi^{\,\ldots BB \ldots}_{\ldots t_j \ldots}$ obtained from such experiments are compared with the co-interference term $\Phi^{\,\ldots AB \ldots}_{\ldots t_j \ldots}$, it is found that they satisfy the following relation: 
\begin{align*}
    &|\Phi^{\,\ldots AB \ldots}_{\ldots t_j \ldots}
            (\ldots a_j^+\!\wedge b_j^+\ldots\,;\,\ldots a_j^-\!\wedge b_j^-\ldots)|\\
        &\phantom{=}= \sqrt{|\Phi^{\,\ldots AA \ldots}_{\ldots t_j \ldots}
            (\ldots a_j^+\!\wedge a_j^+\ldots\,;\,\ldots a_j^-\!\wedge a_j^-\ldots)|}
        \sqrt{|\Phi^{\,\ldots BB \ldots}_{\ldots t_j \ldots}
            (\ldots b_j^+\!\wedge b_j^+\ldots\,;\,\ldots b_j^-\!\wedge b_j^-\ldots)|}.
\end{align*}
This shows that, for independent subsystems, the co-interference term indeed factorizes into a product of subsystem distributions. 

Moreover, it is found that the co-interference term for a system composed of two identical subsystems has the following property:
\begin{align*}
    0 \geq \Phi^{\,\ldots FF \ldots}_{\ldots t_j \ldots}(
        \ldots f_j^+\!\wedge f_j^+\ldots\,;\,\ldots f_j^-\!\wedge f_j^-\ldots
    ) 
    =
    -\Phi^{\,\ldots FF \ldots}_{\ldots t_j \ldots}(
        \ldots f_j^+\!\wedge f_j^-\ldots\,;\,\ldots f_j^-\!\wedge f_j^+\ldots
        ).
\end{align*}
When the subsystems are identical, the co-interference term must have the form of a \textit{square} of a subsystem distribution. Therefore, the above relation indicates that this distribution is a purely imaginary (the square is negative) and anti-symmetric function of the sequence pairs.
\separator
\end{observation}

To account for observations~\ref{obs:which-subsys_interference} and~\ref{obs:identical_subsystems}, the interference term-based parameterization of probability distributions postulated in the previous section, cf.~Eq.~\eqref{eq:diagonal_interference_term}, has to be further expanded. The most straightforward choice is to introduce a complex-valued distributions of sequence pairs \cite{Diosi_PRL2004}, $Q^{\tup{F}}_{\tup{t}}:\Omega(\bm F_n)\times\Omega(\bm F_n) \to \mathbb{C}$, that are hermitian functions with respect to exchange of their arguments $\tup{f}^+\leftrightarrow \tup{f}^-$
\begin{align}\label{eq:bi-prob:hermitianity}
    Q^{\tup{F}}_{\tup{t}}(\tup{f}^-,\tup{f}^+) = Q^{\tup{F}}_{\tup{t}}(\tup{f}^+,\tup{f}^-)^*,    
\end{align}
and exhibits additivity with respect to the coarse-graining operation $\vee$:
\begin{align}
    Q^{\fbarsb{\bm{F}}{n}}_{\tup{t}}\Big(\bigvee_{\omega(\fbarsb{\bm{f}}{n})}\tup{f}^+,\bigvee_{\omega(\fbarsb{\bm{f}}{n})}\tup{f}^-\Big)
        = \sum_{\tup{f}^\pm\in\omega(\fbarsb{\bm{f}}{n})} Q_{\tup{t}}^{\tup{F}}(\tup{f}^+,\tup{f}^-).
\end{align}
By definition, we set their real parts to be equal to the corresponding interference term,
\begin{align}
    I_{\tup{t}}^{\tup{F}}(\tup{f}^+,\tup{f}^-) \equiv 
        \operatorname{Re}Q^{\tup{F}}_{\tup{t}}(\tup{f}^+,\tup{f}^-),
\end{align}
and their imaginary part to be set by the factorization of the co-interference term:
\begin{align}
\nonumber
    \Phi^{\,\ldots AB \ldots}_{\ldots t_j \ldots}(
            \ldots a^+_j\!\wedge b_j^+ \ldots\,;\,\ldots a^-_j\!\wedge b_j^- \ldots
        ) &\equiv \mathrm{i}\operatorname{Im}Q^{\,\ldots A \ldots}_{\ldots t_j \ldots}(
            \ldots a_j^+\ldots\,;\,\ldots a_j^- \ldots
            )\\
        &\phantom{=}\times
            \mathrm{i}\operatorname{Im}Q^{\,\ldots B \ldots}_{\ldots t_j \ldots}(
            \ldots b_j^+\ldots\,;\,\ldots b_j^- \ldots
            );\\
    \sqrt{|\Phi^{\,\ldots FF \ldots}_{\ldots t_j \ldots}(
        \ldots f_j^+\!\wedge f_j^+\ldots\,;\,\ldots f_j^-\!\wedge f_j^- \ldots
    )|} &= 
        |\operatorname{Im}Q^{\,\ldots F \ldots}_{\ldots t_j \ldots}(\ldots f_j^+ \ldots\,;\,\ldots f_j^-\ldots)|.
\end{align}
As the function $Q^{\tup{F}}_{\tup{t}}$ is hermitian and its real part is symmetric, the imaginary part is thus anti-symmetric, which is consistent with observation~\ref{obs:identical_subsystems}.

We can now use this parameterization to rewrite the ``which subsystem'' interference term as a combination of products of distributions associated with independent subsystems,
\begin{align}
\nonumber
&P^{\,\ldots\fbar{AB}\ldots}_{\ldots t_j \ldots}\Big(
    \ldots\bigvee_{\omega\left((\overline{a\wedge b})_j\right)}a_j\wedge b_j \ldots
\Big)\\
\nonumber
    &= I^{\,\ldots \fbar{AB}\ldots}_{\ldots t_j\ldots}\Big(
        \ldots\bigvee_{\omega\left((\overline{a\wedge b})_j\right)}a^+_j\!\wedge b^+_j \ldots\,;\,
        \ldots\bigvee_{\omega\left((\overline{a\wedge b})_j\right)}a^-_j\!\wedge b^-_j \ldots
    \Big)\\
\nonumber
    &= \sum_{(a_j^\pm,b_j^\pm)\in\omega\left((\overline{a\wedge b})_j\right)}
        I^{\,\ldots AB \ldots}_{\ldots t_j \ldots}(
            \ldots a_j^+\!\wedge b_j^+\ldots\,;\,\ldots a_j^-\!\wedge b_j^-\ldots
        )\\
\nonumber
    &=\sum_{(a_j^\pm,b_j^\pm)\in\omega\left((\overline{a\wedge b})_j\right)}\Bigg[
        I^{\,\ldots A \ldots}_{\ldots t_j\ldots}(\ldots a_j^+ \ldots\,;\,\ldots a_j^- \ldots)
        I^{\,\ldots B \ldots}_{\ldots t_j\ldots}(\ldots b_j^+ \ldots\,;\,\ldots b_j^- \ldots)
        \\
\nonumber
    &\phantom{==\sum_{(a_j^\pm,b_j^\pm)\in\omega\left((\overline{a\wedge b})_j\right)}}
    +\Phi^{\,\ldots AB \ldots}_{\ldots t_j\ldots}(
            \ldots a_j^+\!\wedge b_j^+\ldots\,;\,\ldots a_j^-\!\wedge b_j^-\ldots
        )\Big]\\
\nonumber
    &=\sum_{(a_j^\pm,b_j^\pm)\in\omega\left((\overline{a\wedge b})_j\right)}\Bigg[
        \operatorname{Re}Q^{\,\ldots A \ldots}_{\ldots t_j\ldots}(\ldots a_j^+ \ldots\,;\,\ldots a_j^- \ldots)
        \operatorname{Re}Q^{\,\ldots B \ldots}_{\ldots t_j\ldots}(\ldots b_j^+ \ldots\,;\,\ldots b_j^- \ldots)\\
\nonumber
    &\phantom{==\sum_{(a_j^\pm,b_j^\pm)\in\omega\left((\overline{a\wedge b})_j\right)}}
    -\operatorname{Im}Q^{\,\ldots A \ldots}_{\ldots t_j\ldots}(\ldots a_j^+ \ldots\,;\,\ldots a_j^- \ldots)
        \operatorname{Im}Q^{\,\ldots B \ldots}_{\ldots t_j\ldots}(\ldots b_j^+ \ldots\,;\,\ldots b_j^- \ldots)
    \Bigg]\\
    &=\sum_{(a_j^\pm,b_j^\pm)\in\omega\left((\overline{a\wedge b})_j\right)}
        \operatorname{Re}\big\{
        Q^{\,\ldots A \ldots}_{\ldots t_j \ldots}(
            \ldots a_j^+ \ldots\,;\,\ldots a_j^-\ldots)
        Q^{\,\ldots B \ldots}_{\ldots t_j \ldots}(
            \ldots b_j^+ \ldots\,;\,\ldots b_j^-\ldots)
        \big\}.
\end{align}
On the other hand, since the imaginary part is anti-symmetric we have
\begin{align}
    Q^{\tup{F}}_{\tup{t}}(\tup{f},\tup{f}) = \operatorname{Re}Q^{\tup{F}}_{\tup{t}}(\tup{f},\tup{f}) = I^{\tup{F}}_{\tup{t}}(\tup{f},\tup{f}) = P^{\tup{F}}_{\tup{t}}(\tup{f}),
\end{align}
and thus
\begin{align}
\nonumber
&P^{\,\ldots\fbar{AB}\ldots}_{\ldots t_j \ldots}\Big(
    \ldots\bigvee_{\omega\left((\overline{a\wedge b})_j\right)}a_j\wedge b_j \ldots
\Big)\\
\nonumber
    &=Q_{\ldots t_j \ldots}^{\,\ldots \fbar{AB} \ldots}\Big(
        \ldots\bigvee_{\omega\left((\overline{a\wedge b})_j\right)}a^+_j\!\wedge b^+_j \ldots\,;\,
        \ldots\bigvee_{\omega\left((\overline{a\wedge b})_j\right)}a^-_j\!\wedge b^-_j \ldots
    \Big)\\
\nonumber
    &=\sum_{(a_j^\pm,b_j^\pm)\in\omega\left((\overline{a\wedge b})_j\right)}
        Q^{\,\ldots AB \ldots}_{\ldots t_j \ldots}(
            \ldots a_j^+\!\wedge b_j^+ \ldots\,;\,\ldots a_j^-\!\wedge b_j^-\ldots)\\
    &=\sum_{(a_j^\pm,b_j^\pm)\in\omega\left((\overline{a\wedge b})_j\right)}
        \operatorname{Re}
        Q^{\,\ldots AB \ldots}_{\ldots t_j \ldots}(
            \ldots a_j^+\!\wedge b_j^+ \ldots\,;\,\ldots a_j^-\!\wedge b_j^-\ldots),
\end{align}
and the last equality is, again, due to the anti-symmetry of the imaginary part. Comparing the two decompositions of $P^{\,\ldots \fbar{AB}\ldots}_{\ldots t_j\ldots}$, we find the following crucial result: the distributions $Q^{AB}_{\tup{t}}$ obey the factorization law when $A$ and $B$ are independent,
\begin{align}\label{eq:Q_factorization}
    Q^{AB}_{\tup{t}}(\tup{a}^+\!\wedge\tup{b}^+ , \tup{a}^-\!\wedge\tup{b}^-)
        = Q^A_{\tup{t}}(\tup{a}^+,\tup{a}^-)Q^B_{\tup{t}}(\tup{b}^+,\tup{b}^-).
\end{align}
In this way, we recover the factorization property for independent systems, while retaining the ability to parameterize the quantum interference effects in partitionable systems.

\subsection{Markovianity and the initialization events}\label{sec:markovianity}

We shall now turn our attention to the special case of \textit{perfectly fine-grained} devices---that is, devices that are not a coarse-grained variant of any other measuring device; we shall reserve symbols $K,L,\ldots$ for such devices, while leaving $F,G,\ldots$ to indicate generic devices that could be of any degree of graining.

\begin{observation}[Markovianity]\label{obs:markov}
The probability distributions describing measurements with a perfectly fine-grained $K_j$-devices factorize as follows:
\begin{align}\label{eq:markovianity}
    P^{\bm K_n}_{\tup{t}}(\tup{k}) &= P^{K_n|K_{n-1}}_{t_n|t_{n-1}}(k_n|k_{n-1})\cdots P^{K_2|K_1}_{t_2|t_1}(k_2|k_1)P^{K_1}_{t_1}(k_1),
\end{align}
where the factors $P^{K_j|K_{j-1}}_{t_j|t_{j-1}}(k_j|k_{j-1})$ are the conditional probabilities~\eqref{eq:conditional_prob}.
\separator
\end{observation}

In probability theory, the property~\eqref{eq:markovianity} is called \textit{Markovianity} and is traditionally interpreted as a ``lack of memory'' because
\begin{align}\label{eq:memoryless}
    P^{K|\bm K_n}_{t|\bm t_n}(k|\bm k_n) &= 
    \frac{P^{K|K_n}_{t|t_n}(k|k_n)P^{K_n|K_{n-1}}_{t_n|t_{n-1}}(k_n|k_{n-1})\cdots P^{K_2|K_1}_{t_2|t_1}(k_2|k_1)P^{K_1}_{t_1}(k_1)}
    {P^{K_n|K_{n-1}}_{t_n|t_{n-1}}(k_n|k_{n-1})\cdots P^{K_2|K_1}_{t_2|t_1}(k_2|k_1)P^{K_1}_{t_1}(k_1)}
    = P^{K|K_n}_{t|t_n}(k|k_n),
\end{align}
that is, the statistics of the result depend only on the latest prior readout, not on the whole history of all previous readouts.\footnote{
We remark that the ``lack of memory'' should not be confused with ``short memory''. Indeed, ``short memory'' suggests that the statistics of the readout at $t$ would depend only on previous events that occurred in the immediate past---i.e., the previous readouts have impact only when $|t-t_i|$ are in some sense small, while those readouts for which $|t-t_i|$ is large can be disregarded. This is not the case here, as the timings in~\eqref{eq:memoryless} form an arbitrary chronological sequence, so it does not matter whether $t-t_n$ is large or small.
}

It is important to reiterate that Markovianity is inherently linked with the fine-graining of the measurement device. One can easily verify that the property is lost when the deployed devices have less than perfect definition, that is, when a coarse-grained $\fbar{K}$-device is deployed in place of the $K$-device, e.g.,
\begin{align}
\nonumber
    P^{K,\fbar K}_{t_2,t_1}(k_2,k_1^+\vee k_1^-) &= P^K_{t_2,t_1}(k_2,k_1^+) + P^K_{t_2,t_1}(k_2,k_1^-) + 2 \operatorname{Re}Q^K_{t_2,t_1}(k_2,k_2;k_1^+,k_1^-)\\
\nonumber
    &= P^{K|K}_{t_2|t_1}(k_2|k_1^+)P^K_{t_1}(k_1^+) + P^{K|K}_{t_2|t_1}(k_2|k_1^-)P^K_{t_1}(k_1^-) + 2\operatorname{Re}Q^K_{t_2,t_1}(k_2,k_2;k_1^+,k_1^-)\\
    &\neq P^{K|\fbar K}_{t_2|t_1}(k_2|k_1^+\vee k_1^-)P^{\fbar K}_{t_1}(k_1^+\vee k_1^-).
\end{align}
This is in line with the traditional notion of Markovianity: the property is known to be destroyed by coarse-graining~\cite{vanKampen_book}.

So far we have avoided drawing the attention to the fact that the description in terms of probability distributions for the sequences of readouts characterized in Section~\ref{sec:sequence_measure} is not yet complete. The missing piece is the control over the \textit{initial condition} of the experiment; this is, of course, a crucial element that allows the experimenter to reset the physical ``state'' of the system, so that the results of the subsequent measurements can be meaningfully compared with the results obtained in previous (and future) runs of the experiment. Therefore, all multitime probability distributions we have been considering so far should actually be interpreted as probabilities conditioned by some form of ``initialization event'',
\begin{align}
    P^{\bm F_n}_{\tup{t}}(\tup{f})\ \to\  P^{\bm F_n}_{\tup{t}|0}(\tup{f}|\,\text{initialization}\,)    
\end{align}
One could imagine many different ways to realize this ``initialization event'' but, thanks to Markovianity, it is possible to define it in the terms of readouts of the measuring device; for example, if we consider the readout $k_0\in\Omega(K)$ of a perfectly fine-grained $K$-device deployed at time $t=0$ as the initialization, then we have  
\begin{align}
    P^{\bm F_n|K}_{\tup{t}|0}(\tup{f}|k_0) 
    &= P_{\tup{t}|0,-s_1,-s_2,\ldots}^{\bm F_n|K,G,H,\ldots}(\tup{f}|k_0,g_{-1},h_{-2},\ldots)
        =P_{\tup{t}|0,-s_1',-s_2',\ldots}^{\bm F_n|K,G',H',\ldots}(\tup{f}|k_0,g_{-1}',h_{-2}',\ldots)
        =\ldots .
\end{align}
Therefore, this $t=0$ readout of the $K$-device does indeed work as an initialization, since everything that happened to the system before that time cannot influence the subsequent measurements. Moreover, the initialization procedure defined in this way can be generalized to include a degree of uncertainty about the initializing measurement,
\begin{align}\label{eq:phenomenology_of_initialization}
    P^{\bm F_n|p}_{\tup{t}|0}(\tup{f}) := \sum_{K}\sum_{k\in\Omega(K)}P_{\tup{t}|0}^{\bm F_n|K}(\tup{f}|k)p^K(k)\quad\text{such that $\sum_{K,k} p^K(k) = 1,\,p^K(k)\geq 0$},
\end{align}
where $p^K(k)$ denotes the probability that the experiment was initialized with a perfectly fine-grained $K$-device that measured the result $k\in\Omega(K)$.

\subsection{Quantum Zeno effect}\label{sec:Zeno}
Consider an experiment in which some perfectly fine-grained $K$-devices are deployed one after the other with a small delay between the measurements. An intuitive expectation is that, in the limit in which the delay between consecutive measurements is negligible, the second readout should be the same as the first. This is indeed what is observed in experiments; one finds
\begin{align}\label{eq:consecutive_measurement}
    \lim_{\Delta t\to 0^+} P^{K|K}_{t+\Delta t|t}(k|k_0) = \delta_{k,k_0}.
\end{align}
Correlating the $K$-device with its coarse-grained variant---the $\fbar K$-device---also yields an intuitive result,
\begin{align}\label{eq:rapid_measurement}
    \lim_{\Delta t\to 0^+}P_{t+\Delta t|t}^{\fbar K|K}(\fbar k|k_0) = \sum_{k\in\omega(\fbar k)}\delta_{k,k_0}.
\end{align}
However, these intuitions are based on the trajectory picture of classical theories: we assume that a trajectory traced by a classical observable is continuous, and thus, its samples taken within a small interval of time have to overlap. Therefore, it is nigh inevitable that quantum observables, for which the trajectory picture is invalid, will subvert expectations based on classical intuitions when one strings a frequent probing into sequences consisting of more than two measurements. Arguably, the most prominent example of such an intuition-defying result is the emergence of the \textit{quantum Zeno effect} in experiments in which the measuring device is deployed with increasing frequency:
\begin{observation}[Quantum Zeno effect]\label{obs:zeno}
   Given a perfectly fine-grained $K$-device and a duration $t$, one observes
\begin{align}\label{eq:Zeno_effect}
    \lim_{n\to\infty} P^{K|K}_{\tup{s}|0}(k_0,\ldots,k_0|k_0) 
        = \lim_{n\to\infty}\prod_{j=0}^{n-1} P^{K|K}_{s_j + \frac{t}{n}|s_j}(k_0|k_0) = 1
    \quad\text{where $s_j = jt/n$}.
\end{align}
That is, the device readouts becomes ``frozen'' in its initial value as the probing frequency tends to infinity. To put it differently, the probability that one measures any changes in the observable decreases as the frequency at which it is measured increases.
\separator
\end{observation}

Note that phenomenological laws like~\eqref{eq:consecutive_measurement} cannot be confirmed in an actual experimental conditions. The deployment of a physical measuring device must take some finite amount of time, so the delay $\Delta t$ between consecutive measurements cannot ever be actually zero. Therefore, the limit $\Delta t\to 0^+$ is to be understood as an idealization where it is assumed that, in general, physical quantities have regular behavior, so that it is possible to extrapolate trends that were actually observed for small, but still finite, $\Delta t$. In formal terms, unless it is stated otherwise, we always assume that any phenomenological function $\varphi(t)$ is bounded and possesses a well-behaved Taylor series expansion, $\varphi(t+\Delta t) = \varphi(t) + \varphi'(t)\Delta t + (1/2)\varphi''(t)\Delta t^2 + O(\Delta t^3)$. 

At this point it is useful to briefly step outside the phenomenological perspective and comment on the standard-formalism interpretation of the Zeno effect. Within the standard formulation, the use of perfectly fine-grained observable with finite spectrum $|\Omega(K)|=d$ corresponds to a system with a finite-dimensional Hilbert space, while the assumption of regular behavior of phenomenological quantities implies bounded system's Hamiltonian operator. In such circumstances, the Zeno effect may be regarded as generic. Nevertheless, it is known that the effect can fail to materialize when the Hamiltonian is unbounded, because of domain issues preventing a quadratic decay at small times. An exhaustive account of various mathematical and physical conditions for the occurrence of Zeno effect can be found in \cite{Facchi_2008}. Having made this observation, we now return to the phenomenological viewpoint.

With these clarifications out of the way, we find that, when combined with Markovianity, the occurrence of Zeno effect provides some insight regarding the laws that govern the dynamics of the measured observable. In particular, assuming the aforementioned regularity, we can conclude from observation~\ref{obs:zeno} that the \textit{survival probability} must have the following short-time behavior:
\begin{align}\label{eq:surv_short-time}
    P^{K|K}_{t+\Delta t|t}(k|k) &\xrightarrow{\Delta t\to 0^+} 1 - v(k,t)^2\Delta t^2 + O(\Delta t^3).
\end{align}
Indeed, if the linear term did not vanish, then the Zeno effect would be impossible,
\begin{align}
    \prod_{j=0}^{n-1}P^{K|K}_{s_j + \frac{t}{n}|s_j}(k_0|k_0) 
    & = \prod_{j=0}^{n-1}\left(1 + L(k_0,s_j)\frac{t}{n} + O(n^{-2}) \right)
    \xrightarrow{n\to\infty} \mathrm{e}^{\int_0^t L(k_0,s)\mathrm{d}s} \neq 1.
\end{align}

It is important to remark that the Zeno effect is a quintessential quantum (or, in any case, non-classical) effect, as it is incompatible with the trajectory picture. By definition, the measurements of a classical observable $K^\mathrm{cl}$ are described with consistent multi-time probabilities, i.e.,
\begin{align}
\nonumber
    P_{\tup{t}}^{K^\mathrm{cl}}(\tup{k}) 
    &= \sum_{\kappa\in\Omega(K^\mathrm{cl})}P^{K^\mathrm{cl}}_{t_n,\ldots,t_j,s,t_{j-1},\ldots,t_1}(k_n,\ldots,k_j,\kappa,k_{j-1},\ldots,k_1)\\
    &=\sum_{\kappa,\kappa'\in\Omega(K^\mathrm{cl})}P^{K^\mathrm{cl}}_{\ldots,t_j,s,t_{j-1},\ldots,t_{j'},s',t_{j'-1},\ldots}
        (\ldots,k_j,\kappa,k_{j-1},\ldots,k_{j'},\kappa',k_{j'-1},\ldots)
        = \ldots .
\end{align}
If we were to assume that $K^\mathrm{cl}$ exhibits the Zeno effect, then, by using consistency, we would conclude that, for any sequence of measurements, one gets
\begin{align}
    P^{K^\mathrm{cl}}_{\tup{t}|0}(\tup{k}|k_0) &= \lim_{N\to\infty}\sum_{\kappa_j:j\notin\iota_N(\{1,\ldots,n\})}P^{K^\mathrm{cl}}_{\bm{s}_N|0}(\bm{\kappa}_N|k_0) = \prod_{j=1}^n\delta_{k_j,k_0},
\end{align}
with $\iota_N:\{1,\ldots,n\}\to\{1,\ldots,N\}$ such that $s_{\iota(j)} = t_j$ and $\kappa_{\iota(j)} = k_j$ for $j=1,\ldots,n$. Therefore, in classical theories only trivial observables---that is, those that are ``frozen'' because they have no dynamics in the first place---can behave as if they exhibited Zeno effect.

\subsection{Multiple observables and uncertainty relations}\label{sec:uncertinty_relations}

We previously noticed (see Section~\ref{sec:sequence_measure}) that, due to the trajectory picture being in effect, any measurable quantity of a classical system can be formally represented as a function of a trajectory representing the elementary system observable. Generally, the elementary observable has the form of a vector-valued trajectory of dimension equal to the number of degrees of freedom, e.g., a position in the three-dimensional space, $E^\mathrm{cl}: t\mapsto (r_x(t),r_y(t),r_z(t))\in\mathbb{R}^3$. Let us consider here a simple case of classical system whose elementary observable is one-dimensional and discrete-valued, $E^\mathrm{cl}: t\mapsto e(t)\in\Omega(E^\mathrm{cl})$ where $\Omega(E^\mathrm{cl})$ is a finite set. Then, an observable $K^\mathrm{cl}: t\mapsto k(t) = K(e(t))\in\Omega(K^\mathrm{cl})$ can be considered as a classical analogue of a perfectly fine-grained observable provided that the function $K: \Omega(E^\mathrm{cl})\to\Omega(K^\mathrm{cl})$ is a bijection (one-to-one function); in that case, measurements of $K^\mathrm{cl}$ are essentially equivalent to measurements of the elementary observable itself,
\begin{align}
\nonumber
    P^{K^{\mathrm{cl}}}_{\tup{t}}(\tup{k}) &= \int \Big(\prod_{j=1}^n p^{K^\mathrm{cl}}_{t_j}\big(k_j\big|e(t_j)\big)\Big)\mathcal{P}^{E^\mathrm{cl}}[\,e\,][\mathcal{D}e]
        = \int \Big(\prod_{j=1}^n \delta_{k_j,K(e(t_j))}\Big)\mathcal{P}^{E^\mathrm{cl}}[\,e\,][\mathcal{D}e]\\
    & = P^{E^{\mathrm{cl}}}_{t_n,\ldots,t_1}\big(K^{-1}(k_n),\ldots,K^{-1}(k_1)\big) = P^{E^\mathrm{cl}}_{\tup{t}}\big(K^{-1}(\tup{k})\big).
\end{align}
Therefore, in classical theories there is little incentive to ever make any distinctions between observables; ultimately, all fine-grained measurements yield essentially the same information. We can summarize this observation with the following \textit{certainty relations},
\begin{align}
\nonumber
        \delta_{e,e_0} &=\lim_{\Delta t\to 0^+} P^{E^\mathrm{cl}|E^\mathrm{cl}}_{t+\Delta t|t}(e|e_0)\\
\nonumber
        &=\lim_{\Delta t\to 0^+} P^{K^\mathrm{cl}|K^\mathrm{cl}}_{t+\Delta t|t}(K(e)|K(e_0))
            = \lim_{\Delta t\to 0^+} P^{L^\mathrm{cl}|L^\mathrm{cl}}_{t+\Delta t|t}(L(e)|L(e_0))\\
        &= \lim_{\Delta t\to 0^+} P^{K^\mathrm{cl}|L^\mathrm{cl}}_{t+\Delta t|t}(K(e)|L(e_0))
            = \lim_{\Delta t\to 0^+} P^{L^\mathrm{cl}|K^\mathrm{cl}}_{t+\Delta t|t}(L(e)|K(e_0)),
\end{align}
where $K^\mathrm{cl}$ and $L^\mathrm{cl}$ are any two perfectly fine-grained observables, and $E^\mathrm{cl}$ is the elementary observable of the classical system.

Since the trajectory picture is invalid for quantum systems, it is possible to have physically inequivalent quantum observables. Consequently, quantum observables are not constrained by the classical certainty relations. So far, we have investigated only a special kind of such inequivalent observables, namely, an arbitrarily chosen observable $F$ and its various (quantum) coarse-grained variants $\fbar{F}$ (cf.~Section~\ref{sec:interference}, Section~\ref{sec:independent_sys}). However, even observables measured with perfectly fine-grained devices can be inequivalent:

\begin{observation}[Uncertainty relations]\label{obs:uncertainty}
Given a pair of perfectly fine-grained devices ($K$- and $L$-device), one observes the following relation when the two devices are deployed in a rapid succession:
\begin{align}\label{eq:uncertinty}
    \lim_{\Delta t\to 0^+} P^{K|L}_{t+\Delta t|t}(k|\ell) 
        = \lim_{\Delta t\to 0^+} P^{L|K}_{t+\Delta t|t}(\ell|k) \equiv C^{K|L}_{k,\ell},\quad\text{for any $t>0$,}
\end{align}
where $C^{K|L}_{k,\ell} = \delta_{k,h(\ell)}$ (with $h:\Omega(L)\to\Omega(K)$ a bijection) when $L=K$, but in general $C^{K|L}_{k,\ell}\neq \delta_{k,h(\ell)}$, i.e., quantum observables can violate certainty relations. 

Moreover, for every perfectly fine-grained $K$-device, it is always possible to find a perfectly fine-grained $K^\perp$-device such that
\begin{align}
    C^{K|K^\perp}_{k,k'} = |\Omega(K)|^{-1},\quad\text{for every $k\in\Omega(K)$ and $k'\in\Omega(K^\perp)$.}
\end{align}
\separator
\end{observation}

Since $C^{K|L}_{k,\ell}$ does not depend on the time at which the measurements were taken, it can serve as a phenomenological measure of how ``similar'' the observables measured by the $K$- and $L$-devices are. On one extreme we have $C^{K|L}_{k,\ell} = \delta_{k,h(\ell)}$, signifying that $K$ and $L$ are equivalent (analogously to all classical observables), while on the other end of the spectrum we have $C^{K|L}_{k,\ell} = |\Omega(K)|^{-1}$, meaning that there is no correlation whatsoever between observations of $L$ and $K$. 

It is important to underline that for classical observers there is no intuitive basis for how to interpret cases $C^{K|L}_{k,\ell} \neq \delta_{k,h(\ell)}$. Indeed, since in classical physics there is essentially only one observable (and thus certainty relations are always true), it is impossible to extrapolate from the classical scenario to any situation in which there are two or more inequivalent observables. Any understanding of measurement, observation etc. acquired in classical setting has the existence of a single trajectory and the equivalence of all observables as foundational principles; there is simply no way to imagine how such a picture would be modified when observables are allowed to be inequivalent.

\section{The bi-trajectory picture of quantum mechanics}\label{sec:bi-traj_picture}

\subsection{Basis for the deduced formalism}
The observations~\ref{obs:pair-wise_interference},~\ref{obs:independent_subsystems} and~\ref{obs:identical_subsystems} concerning various aspects of quantum interference described in Section~\ref{sec:phenomenology} compelled us to consider a formal parameterization of phenomenological probability distributions based on a family of complex-valued distributions on the space of sequence pairs,
\begin{align}
    Q^{\tup{F}}_{\tup{t}}:\Omega(\bm F_n)\times\Omega(\bm F_n) \to \mathbb{C}.
\end{align}
These distributions allowed us to succinctly describe the phenomenology of interference effects, and to conveniently encode the crucial observation that interference occurs only between pairs of alternatives within the very structure of the functions employed,
\begin{align}\label{eq:pair-wise_decomposition}
    P^{\fbarsb{\bm{F}}{n}}_{\tup{t}}(\fbarsb{\bm f}{n})
        = Q^{\fbarsb{\bm{F}}{n}}_{\tup{t}}(\fbarsb{\bm{f}}{n},\fbarsb{\bm{f}}{n}) 
        = \sum_{\tup{f}^\pm\in\omega(\fbarsb{\bm f}{n})}
            Q^{\tup{F}}_{\tup{t}}(\tup{f}^+,\tup{f}^-)
        = \sum_{\tup{f}^\pm\in\omega(\fbarsb{\bm f}{n})}
            \operatorname{Re}Q^{\tup{F}}_{\tup{t}}(\tup{f}^+,\tup{f}^-).
\end{align}
Furthermore, through a consistent application of the properties we have defined the functions $Q_{\tup{t}}^{\tup{F}}$ to possess, we proved that they obey the factorization rule for systems partitionable into independent subsystems [cf.~Eq.~\eqref{eq:Q_factorization}]. This is a key feature, without which it would be impossible to define a coherent formal description of isolated systems (i.e., a system independent of any other system), and thus, any possibility of formulating a functioning formalism for the theory would be negated.

Consequently, the clear advantages of the distributions $Q^{\tup{F}}_{\tup{t}}$ makes them a prime candidate as the basis for the formalism that we strive to deduce from the phenomenological observations. However, concerns remain regarding whether the choice of $Q^{\tup{F}}_{\tup{t}}$ as the formal parameterization ingredient conflicts with the ethos of phenomenological quantum mechanics. To this regard we notice that, for the class of experiments analyzed so far, only a limited number of values $Q_{\tup{t}}^{\tup{F}}(\tup{f}^+,\tup{f}^-)$ can be considered as phenomenological quantities. These are (a) the diagonal values $Q^{\tup{F}}_{\tup{t}}(\tup{f},\tup{f})$, because they simply equal the measurable probabilities, and (b) the interference terms involved in the pair-wise decomposition in a \textit{single} $k$-slit experiment,
\begin{align}
\nonumber
    &\operatorname{Re}Q^{\tup{F}}_{\tup{t}}(f_n,\ldots,f_j^+,\ldots,f_1\,;\,f_n,\ldots,f_j^-,\ldots,f_1)\\
    &\phantom{=}=\frac{1}{2}\Big[P^{F_n,\ldots,\fbarsb{F}{j},\ldots,F_1}_{t_n,\ldots,t_j,\ldots,t_1}(f_n,\ldots,f_j^+\vee f_j^-,\ldots,f_1)
        - P^{\tup{F}}_{\tup{t}}(\ldots,f_j^+,\ldots)
        - P^{\tup{F}}_{\tup{t}}(\ldots,f_j^-,\ldots)\Big].
\end{align}

On the other hand, as an alternative choice to distributions $Q_{\tup{t}}^{\tup{F}}$, which are generally non-phenomenological, we might consider functions defined as their symmetrized versions; namely,
\begin{align}\label{eq:sym_I}
    \mathcal{S}[Q^{\tup{F}}_{\tup{t}}](f_n^+,f_n^-;\ldots;f_1^+,f_1^-) 
        := \prod_{j=1}^n\Bigg(\sum_{(\phi_j^+,\phi_j^-)\in S(\{f_j^+,f_j^-\})}\Bigg)
            Q^{\tup{F}}_{\tup{t}}(
                \phi_n^+,\ldots,\phi_1^+\,;\,\phi_n^-,\ldots,\phi_1^-),
\end{align}
where $S(A)$ denotes the set of all permutations of elements of $A$, for example $S(\{f_j^+,f_j^-\}) = \{ (f_j^+,f_j^-), (f_j^-,f_j^+) \}$ and $S(\{f_j,f_j\})=\{(f_j,f_j)\}$. All such functions decompose into combinations of phenomenological probability distributions, thus making them phenomenological quantities; for example
\begin{align}
\nonumber
    &\mathcal{S}[Q^{\tup{F}}_{\tup{t}}](f_n,f_n;\ldots;f_j^+,f^-_j;\ldots;f_k^+,f_k^-;\ldots;f_1,f_1)\\
\nonumber
    &\phantom{=}=2\operatorname{Re}Q^{\tup{F}}_{\tup{t}}(\ldots f_j^+ \ldots f_k^+ \ldots \,;\, \ldots f_j^- \ldots f_k^- \ldots) 
    + 2\operatorname{Re}Q^{\tup{F}}_{\tup{t}}(\ldots f_j^+ \ldots f_k^- \ldots \,;\, \ldots f_j^- \ldots f_k^+ \ldots)\\
\nonumber
    &\phantom{=}=
        P^{\,\ldots\fbarsb{F}{j}\ldots\fbarsb{F}{k}\ldots}_{\ldots t_j \ldots t_k \ldots}(\ldots f_j^+\vee f_j^- \ldots f_k^+\vee f_k^- \ldots)\\
\nonumber
    &\phantom{====}
        - P^{\,\ldots \fbarsb{F}{j}\ldots F_k \ldots}_{\ldots t_j \ldots t_k \ldots}(\ldots f_j^+\vee f_j^- \ldots f_k^+ \ldots)
        - P^{\,\ldots \fbarsb{F}{j}\ldots F_k \ldots}_{\ldots t_j \ldots t_k \ldots}(\ldots f_j^+\vee f_j^- \ldots f_k^- \ldots)\\
\nonumber
    &\phantom{====}
        - P^{\,\ldots F_j \ldots \fbarsb{F}{k} \ldots}_{\ldots t_j \ldots t_k \ldots}(\ldots f_j^+ \ldots f_k^+\vee f_k^- \ldots)
        - P^{\,\ldots F_j\ldots \fbarsb{F}{k} \ldots}_{\ldots t_j \ldots t_k \ldots}(\ldots f_j^- \ldots f_k^+\vee f_k^- \ldots)\\
\nonumber
    &\phantom{====}
        + P^{\,\ldots F_j \ldots F_k \ldots}_{\ldots t_j \ldots t_k \ldots}(\ldots f_j^+ \ldots f_k^+ \ldots)
        + P^{\,\ldots F_j\ldots F_k \ldots}_{\ldots t_j \ldots t_k \ldots}(\ldots f_j^- \ldots f_k^-\ldots)\\
&\phantom{====}
        + P^{\,\ldots F_j \ldots F_k \ldots}_{\ldots t_j \ldots t_k \ldots}(\ldots f_j^+ \ldots f_k^- \ldots)
        + P^{\,\ldots F_j\ldots F_k \ldots}_{\ldots t_j \ldots t_k \ldots}(\ldots f_j^- \ldots f_k^+ \ldots)
    .
\end{align}
This raises the following objection: perhaps, it would be more sensible to simply use $\mathcal{S}[Q^{\tup{F}}_{\tup{t}}]$ as the formal parameterization, rather than the ``bare'' functions $Q_{\tup{t}}^{\tup{F}}$. After all, the pair-wise interference decomposition in observation~\ref{obs:pair-wise_interference} only consists of symmetric combinations anyway,
\begin{align}
\nonumber
    P^{\fbarsb{\bm{F}}{n}}_{\tup{t}}\Big(\bigvee_{\omega(\fbarsb{\bm{f}}{n})}\tup{f}\Big)
    &= \sum_{\tup{f}^\pm\in\omega(\fbarsb{\bm{f}}{n})} Q^{\tup{F}}_{\tup{t}}(\tup{f}^+,\tup{f}^-)
        = \prod_{j=1}^n\Bigg(\sum_{(f_j^+,f_j^-)\in\omega(\fbarsb{f}{j})\times\omega(\fbarsb{f}{j})}\Bigg)
        Q^{\tup{F}}_{\tup{t}}(\tup{f}^+; \tup{f}^-)\\
\nonumber
    &= \prod_{j=1}^n\Bigg(
        \sum_{\substack{\{f_j^+,f_j^-\}\mid f_j^\pm\in\omega(\fbarsb{f}{j})}}\,
        \sum_{(\phi_j^+,\phi_j^-)\in S(\{f_j^+,f_j^-\})}\Bigg)
            Q^{\tup{F}}_{\tup{t}}(
                \phi_n^+,\ldots,\phi_1^+\,;\,\phi_n^-,\ldots,\phi_1^-
            )\\
    &= \prod_{j=1}^n\Bigg(\sum_{\{f_j^+,f_j^-\}|f_j^\pm\in\omega(\fbarsb{f}{j})}\Bigg)
        \mathcal{S}[Q_{\tup{t}}^{\tup{F}}](f_n^+,f_n^-;\ldots;f_1^+,f_1^-).
\end{align}
Hence, when debating the introduction of an \textit{ad hoc} formal element, what standard should be used to determine whether distributions like $Q_{\tup{t}}^{\tup{F}}$ should be included over alternatives such as $\mathcal{S}[Q_{\tup{t}}^{\tup{F}}]$? In response, we propose the following two criteria:
\begin{enumerate}[label=(\roman*)]
    \item There must be a compelling case demonstrating why the purpose of this new element cannot be fulfilled by phenomenological elements (the probability distributions $P^{\tup{F}}_{\tup{t}}$ and $\mathcal{S}[Q_{\tup{t}}^{\tup{F}}]$ in this case). 
    \item Even if such a case is successfully made, minimal extensions of the phenomenological elements should be preferred over formulations that introduce additional, potentially excessive, complexity. 
\end{enumerate}
We believe that the distributions $Q_{\tup{t}}^{\tup{F}}$ comply with both requirements.

First, the necessity for introducing an extension from single-sequence to sequence-pair distributions is evident when considering the difficulty of clearly expressing the crucial pair-wise interference decomposition by only using the phenomenological probability distributions. This issue alone does not present a strong enough case for adopting the ``bare'' $Q_{\tup{t}}^{\tup{F}}$, since this decomposition can also be accomplished using the phenomenological $\mathcal{S}[Q^{\tup{F}}_{\tup{t}}]$. However, the case for $Q_{\tup{t}}^{\tup{F}}$ is further strengthened by the demonstration of the factorization rule in ``which subsystem'' interference experiments---something that would remain obscured if only probabilities were used. Moreover, the case of ``which subsystem'' interference effects also disqualifies $\mathcal{S}[Q_{\tup{t}}^{\tup{F}}]$, as the symmetrization eliminates the anti-symmetric imaginary part of $Q_{\tup{t}}^{\tup{F}}$, which is crucial for parameterizing the observed co-interference term.

Second, the distributions $Q_{\tup{t}}^{\tup{F}}$ can be reasonably considered the simplest possible extension of phenomenological probabilities. Transitioning from $P^{\tup{F}}_{\tup{t}}$, which are (probability) distributions on the space of sequences $\tup{f}\in\Omega(\tup{F})$, to $Q_{\tup{t}}^{\tup{F}}$, which are (non-probability) distributions on the space of sequence pairs $(\tup{f}^+,\tup{f}^-)\in\Omega(\bm F_n)\times\Omega(\bm F_n)$, is the simplest and most natural progression of the formal description. In contrast, due to their symmetries, the functions $\mathcal{S}[Q_{\tup{t}}^{\tup{F}}]$ should be viewed as distributions on the space of $n$ pairs of alternatives. In fact, since $\mathcal{S}[Q^{\tup{F}}_{\tup{t}}]$ are defined as symmetrized interference terms (equal to the symmetric real part of $Q_{\tup{t}}^{\tup{F}}$), they should be viewed as functions of, up to, $n$ different sequences---an extreme increase in complexity compared to $P^{\tup{F}}_{\tup{t}}$.

\subsection{Bi-probability distributions}

If we accept the distributions $Q_{\tup{t}}^{\tup{F}}$ as a valid extension of the formal description, the next step is to collate the data from the analyzed experiments and verify that the parameterization with these functions remains consistent with all phenomenological observations made so far. In principle, the most straightforward approach would be to treat each measurement outcome as an \textit{equality constraint}, where the values $Q_{\tup{t}}^{\tup{F}}(\tup{f}^+,\tup{f}^-)$ are the unknowns to be solved for, and the phenomenological probabilities $P_{\tup{t}}^{\tup{F}}(\tup{f})$ are the given parameters. For instance, constraints such as $\operatorname{Re}Q_{\tup{t}}^{\tup{F}}(\tup{f},\tup{f}) - P_{\tup{t}}^{\tup{F}}(\tup{f}) = 0$ and $\sum_{\tup{f}^\pm\in\omega(\fbarsb{f}{n})}Q_{\tup{t}}^{\bm{\bm F}_n}(\tup{f}^+,\tup{f}^-)-P^{\fbarsb{\bm F}{n}}_{\tup{t}}(\fbarsb{\bm f}{n})=0$, would be used. However, we know that such a system of equations does not have a single unique solution---were it so, the distributions $Q_{\tup{t}}^{\tup{F}}$ themselves would be phenomenological quantities! Consequently, in such an approach, the constraints would constitute an infinite number of conditions defining the functions $Q_{\tup{t}}^{\tup{F}}$, which violates the requirement of minimizing the added complexity.

Instead of attempting to solve every constraining equation, we should focus on identifying a \textit{minimal set of properties} for the distributions $Q_{\tup{t}}^{\tup{F}}$ that would be sufficient for them to serve as a viable solutions. When composing such a list, aside from minding the contents of phenomenological observations, we can aid our efforts by considering how the properties of uni-sequence probability distributions could be generalized for the case of complex-valued distributions of bi-sequence. The following properties can be identified in using this approach:
\begin{enumerate}[label=\textnormal{(Q\arabic*)}]
    \item\label{prop:bi-prob:norm} \textit{Normalization}
        \begin{align*}
            \sum_{\tup{f}^\pm\in\Omega(\tup{F})}Q^{\tup{F}}_{\tup{t}}(\tup{f}^+,\tup{f}^-) = 1    
        \end{align*}
    \item\label{prop:bi-prob:causality} \textit{Causality}:
        \begin{align*}
            Q^{F_n,\ldots,F_1}_{t_n,\ldots,t_1}(f_n^+,\ldots, f_1^+\,;\,f_n^-,\ldots,f_1^-) \propto \delta_{f_n^+,f_n^-};
        \end{align*}
    \item\label{prop:bi-prob:factorization}\textit{Factorization rule for independent systems}:
        \begin{align*}
            Q^{AB}_{\tup{t}}(\bm{a}^+_n\!\wedge\tup{b}^+ , \tup{a}^-\!\wedge\tup{b}^-)
                = Q^A_{\tup{t}}(\tup{a}^+,\tup{a}^-)Q^B_{\tup{t}}(\tup{b}^+,\tup{b}^-);
        \end{align*}
    \item\label{prop:bi-prob:pos} \textit{Positive semi-definiteness}:
        \begin{align*}
            \sum_{\tup{f}^\pm\in\Omega(\tup{F})}Z(\tup{f}^+)Q^{\tup{F}}_{\tup{t}}(\tup{f}^+,\tup{f}^-)Z(\tup{f}^-)^* \geq 0\quad\text{for any function $\tup{f}\mapsto Z(\tup{f})\in\mathbb{C}$};
        \end{align*}
        This property can be rephrased as follows: $\hat Q^{\tup{F}}_{\tup{t}} := [Q^{\tup{F}}_{\tup{t}}(\tup{f}^+,\tup{f}^-)]_{\tup{f}^-,\tup{f}^+}$, where sequences $\tup{f}^\pm\in\Omega(\tup{F})$ are treated as indexes, is a positive semi-definite matrix.
    \item\label{prop:bi-prob:bi-consistency} \textit{Bi-consistency}:
        \begin{align*}
            \sum_{f_j^\pm\in\Omega(F_j)}Q^{\tup{F}}_{\tup{t}}(\tup{f}^+,\tup{f}^-) = Q^{F_n,\ldots,\cancel{F_j},\ldots,F_1}_{t_n,\ldots,\cancel{t_j},\ldots,t_1}(f_n^+,\ldots,\cancel{f_j^+},\ldots,f_1^+\,;\,f_n^-\ldots,\cancel{f_j^-},\ldots,f_1^-);
        \end{align*}
\end{enumerate}
Functions $Q_{\tup{t}}^{\bm F_n}$ satisfying the properties~\ref{prop:bi-prob:norm}--\ref{prop:bi-prob:bi-consistency} are referred to as \textit{bi-probability distributions} in~\cite{Szankowski_Quantum24,Lonigro_Quantum24}, and we shall stick to this convention; the name \textit{decoherence functional} can also be found in the literature~\cite{Sorkin_JoPA07,Sorkin_ModPhysLettA94,gudder2009quantum,gudder2010finite,Gudder_MathSlov12}.

Examining the listed properties, we see that~\ref{prop:bi-prob:norm} is a straightforward generalization of the normalization condition for probability distributions. The property~\ref{prop:bi-prob:causality}, for obvious reasons, does not correspond to any property of uni-sequence distributions; instead, given the properties~\ref{prop:bi-prob:norm} and~\ref{prop:bi-prob:bi-consistency}, it is a necessary and sufficient condition for the bi-probabilities $Q_{\tup{t}}^{\bm F_n}$ to be compatible with observations~\ref{obs:causality} and~\ref{obs:coarse-grained_measurement}. 
The factorization rule~\ref{prop:bi-prob:factorization} simply reiterates the findings of Section~\ref{sec:independent_sys}, see Eq.~\eqref{eq:Q_factorization}. The positive semi-definiteness~\ref{prop:bi-prob:pos} can be seen as a generalization of the non-negativity of uni-sequence probability distributions. It is also the ``workhorse'' property: it enforces the Hermicity~\eqref{eq:bi-prob:hermitianity}, and more importantly, when combined with normalization~\ref{prop:bi-prob:norm}, causality~\ref{prop:bi-prob:causality}, and~\ref{prop:bi-prob:bi-consistency}, it implies
\begin{align}
    Q^{\tup{F}}_{\tup{t}}(\tup{f},\tup{f}) 
        = \operatorname{Re}Q_{\tup{t}}^{\bm F_n}(\tup{f},\tup{f}) \geq 0 
        \quad\text{and}\quad
    \sum_{\tup{f}\in\Omega(\tup{F})}Q^{\tup{F}}_{\tup{t}}(\tup{f},\tup{f})= 1.
\end{align}
indicating that the diagonal values of bi-probabilities can serve as uni-sequence multi-time probability distributions, which exhibit both the causality
\begin{align}
    &\sum_{f_{n+1}}Q^{\tup[n+1]{F}}_{\bm t_{n+1}}(\bm f_{n+1},\bm f_{n+1})
        = \sum_{f_{n+1}}\sum_{f_{n+1}'}\delta_{f_{n+1},f'_{n+1}}
            Q^{\tup[n+1]{F}}_{\bm t_{n+1}}(f_{n+1},\bm f_n\,;\,f'_{n+1},\bm f_n)
        = Q_{\tup{t}}^{\bm F_n}(\bm f_n,\bm f_n),
\end{align}
and inconsistency,
\begin{align}
\nonumber
    \ref{prop:bi-prob:bi-consistency}\implies
    Q^{F_n,\ldots,\cancel{F_j},\ldots,F_1}_{t_n,\ldots,\cancel{t_j},\ldots,t_1}(
            f_n,\ldots,\cancel{f_j}\ldots,f_1\,;\,f_n,\ldots,\cancel{f_j},\ldots,f_1)
        - \sum_{f_j\in\Omega(F_j)}Q^{\bm F_n}_{\bm t_n}(\bm f_n,\bm f_n)&\\
\label{prop:bi-prob:inconsistency}
        = \sum_{f_j^+\neq f_j^-}Q_{\tup{t}}^{\bm F_n}(
            f_n,\ldots,f_j^+,\ldots f_1\,; f_n,\ldots,f_j^-,\ldots,f_1)\neq 0.&
\end{align}
Moreover, given a collection of resolutions $\operatorname{Res}(\fbarsb{F}{j}|F_j)$ for $j=1,\ldots,n$ defined in accordance with Eq.~\eqref{eq:resolution_Fbar}, positive semi-definiteness also implies
\begin{align}
    \sum_{\tup{f}^\pm\in\omega(\fbarsb{\bm f}{n})}Q_{\tup{t}}^{\tup{F}}(\tup{f}^+,\tup{f}^-)\geq 0
    \quad\text{and}\quad
    \sum_{\fbarsb{\bm{f}}{n}\in\Omega(\fbarsb{\bm F}{n})}\sum_{\tup{f}^\pm\in\omega(\fbarsb{\bm f}{n})}
        \!\!Q_{\tup{t}}^{\tup{F}}(\tup{f}^+,\tup{f}^-)
    = \!\!\sum_{\tup{f}^\pm\in\Omega(\bm F_n)}\!\!Q^{\bm F_n}_{\bm t_n}(\bm f_n^+, \bm f_n^-) = 1
\end{align}

For these reasons, bi-probability distributions can be used to parameterize the multi-time probabilities compliant with phenomenological observations~\ref{obs:causality} and~\ref{obs:inconsistency}, and the pair-wise interference decomposition~\eqref{eq:pair-wise_decomposition},
\begin{align}\label{prop:bi-prob:F-link}
    P^{\tup{F}|p}_{\tup{t}|t_0}(\tup{f}) &= Q^{\tup{F}|p}_{\tup{t}|t_0}(\tup{f},\tup{f}).
\end{align}
Notice that we can exploit the link with phenomenological probabilities to incorporate into bi-probabilities the information about the initialization event.

Finally, we have the \textit{bi-consistency}~\ref{prop:bi-prob:bi-consistency}. This can be seen as a generalization of the classical Kolmogorov consistency condition (cf.~Section~\ref{sec:coarse-graining}) adopted for distributions on pairs of sequences. It is also a sufficient condition for functions $Q_{\tup{t}}^{\tup{F}}$ to be compliant with observation~\ref{obs:extrm_coarse} of the effects of the extreme coarse-grained measurements. 

\subsection{The bi-trajectory picture}\label{sec:extension_thrm}
As we can see from Eq.~\eqref{prop:bi-prob:inconsistency}, even though bi-probabilities are directly responsible for violating the consistency of the probability distributions $P^{\bm F_n}_{\tup{t}}$ through interference effects---and thus invalidating the trajectory picture for quantum observables---they still retain an analogous notion~\ref{prop:bi-prob:bi-consistency}: rather than for a sequence of results, it is the \textit{pair} of sequences of results that is consistent here. This begs the question if, in analogy to the extension theorem of classical probability theory, bi-consistency implies the existence of a measure on the space of trajectory \textit{pairs}.

As it was discussed in~\cite{Lonigro_Quantum24}, from the mathematical point of view, when non-positive-valued distributions are involved, bi-consistency~\ref{prop:bi-prob:bi-consistency} alone is not sufficient for the bi-trajectory measure to exists. Nevertheless, we can show that the observations~\ref{obs:markov} (Markovianity) and~\ref{obs:zeno} (Zeno effect)---the two phenomenological properties that give us a glimpse into the inner-workings of the quantum dynamical laws---imply that bi-probabilities \textit{do} satisfy the additional necessary condition for the generalized Kolmogorov extension theorem. The formal proof of this assertion goes as follows.

Let $\mathbb{Q}_{[0,T]}^{F|p} = \{Q^{F|p}_{\tup{t}|0}\mid n\in\mathbb{N},t_1,\ldots,t_n\in [0,T], 0<t_1<\cdots<t_n\}$ be the family of bi-probabilities associated with observable $F$ and the finite time window $[0,T]$. It was previously proven---see~\cite[Theorem 2.2]{Lonigro_Quantum24}---that the family $\mathbb{Q}_{[0,T]}^{F|p}$ extends to the bi-trajectory measure $\mathcal{Q}_{[0,T]}^{F|p}\sqm{f}$ provided that (i) it is bi-consistent; and (ii) it is \textit{uniformly bounded},
\begin{align}
    \sup\Big\{ \big\|Q^{F|p}_{\tup{t}|0}\big\|_1 \ \Big|\  Q^{F|p}_{\tup{t}|0}\in\mathbb{Q}_{[0,T]}^{F|p}\Big\} < \infty\qquad 
    \text{where}\quad\big\|Q^{F|p}_{\tup{t}|0}\big\|_1 := \sum_{\tup{f}^\pm}|Q^{F|p}_{\tup{t}|0}(\tup{f}^+,\tup{f}^-)|.
\end{align}
In this case, bi-consistency is automatically satisfied~\ref{prop:bi-prob:bi-consistency}, but uniform boundedness has to be verified: (1) Due to bi-consistency we have (see also~\cite{Lonigro_Quantum24}, Lemma 3.5 and Proposition 3.6)
\begin{align}
    \big\| Q^{F|p}_{\tup{t}|0}\big\|_1 \leq \lim_{n\to\infty}\big\| Q^{F|p}_{\tup{s}}\big\|_1\qquad\text{where}\quad s_j = j T/n.
\end{align}
(2) The property~\ref{prop:bi-prob:pos}, i.e., $Q^{F|p}_{\tup{t}|0}(\tup{f}^+,\tup{f}^-)$ is a positive \textit{matrix} when one treats $\tup{f}^\pm$ as indexes, implies that
\begin{align}
    |Q^{F|p}_{\tup{s}|0}(\tup{f}^+,\tup{f}^-)| &\leq \sqrt{Q^{F|p}_{\tup{s}|0}(\tup{f}^+,\tup{f}^+)Q^{F|p}_{\tup{s}|0}(\tup{f}^-,\tup{f}^-)} 
        = \sqrt{P^{F|p}_{\tup{s}|0}(\tup{f}^+)}\,\sqrt{P^{F|p}_{\tup{s}|0}(\tup{f}^-)}.
\end{align}
(3) The short-time behavior of the survival probability~\eqref{eq:surv_short-time}, implied by the Markov property~\eqref{eq:markovianity} and the Zeno effect, allows us to estimate the square root of probability,
\begin{align}
\nonumber
    \sqrt{P^{F|p}_{\tup{s}|0}(\tup{f})} &= \sqrt{P^{F|p}_{s_1|0}(f_1)}\prod_{j=1}^{n-1} \sqrt{P^{F|F}_{s_{j+1}|s_{j}}(f_{j+1}|f_{j})}\\
\nonumber
    &\leq |\Omega(F)|\prod_{j=1}^{n-1} \bigg(\delta_{f_{j+1},f_{j}}+(1-\delta_{f_{j+1},f_{j}})\sqrt{1-P^{F|F}_{s_{j}+\frac{T}{n}|s_{j}}(f_{j}|f_j)}\bigg)\\
        &\leq |\Omega(F)|\prod_{j=1}^{n-1} \Big(\delta_{f_{j+1},f_{j}}+(1-\delta_{f_{j+1},f_{j}})\Big(\sup_{f}|v(f,s_{j})|\frac{T}{n}+O(n^{-2})\Big)\Big),
\end{align}
which then leads to the required uniform bound,
\begin{align}
\nonumber
    \sup\big\|Q^{F|p}_{\tup{t}|0}\big\|_1 &\leq |\Omega(F)|^2\lim_{n\to\infty}\left[\prod_{j=1}^{n-1}\sum_{f_j}
        \Big(\delta_{f_{j+1},f_{j}}+(1-\delta_{f_{j+1},f_{j}})\Big(\sup_{f}|v(f,s_{j})|\frac{T}{n}+O(n^{-2})\Big)\Big)\right]^2\\
\nonumber
    &\leq |\Omega(F)|^2 \lim_{n\to\infty}\left[\prod_{j=1}^{n-1}\Big(1 + |\Omega(F)|\sup_{f}|v(f,s_j)|\frac{T}{n} + O(n^{-2})\Big)\right]^2\\
    &\leq |\Omega(F)|^2\, {\exp}\Big[2|\Omega(F)|\int_0^T \sup_f|v(f,s)|\mathrm{d}s\Big] < \infty.
\end{align}

Therefore, $\mathcal{Q}_{[0,T]}^{F|p}$ extends the family $\mathbb{Q}_{[0,T]}^{F|p}$ for any choice of $T$. But this particular $\mathcal{Q}^{F|p}_{[0,T]}$ can also be seen as a restriction (to time window $[0,T]$) of a measure $\mathcal{Q}^{F|p}$ that extends $\mathbb{Q}^{F|p} = \lim_{n\to\infty}\mathbb{Q}_{[0,T_n]}^{F|p}$ such that $T_{1} < T_2 < \cdots < T_{n}$; compare with~\cite{Lonigro_Quantum24}, Section 4.1. This concludes the proof and leads us to the conclusion that it is possible to deduce at this point that the bi-trajectory measure does, indeed, exist:
\begin{enumerate}[label=\textnormal{(Q\arabic*)},resume]
    \item\label{prop:bi-prob:master_measure} \textit{The bi-trajectory measure}: The bi-probabilities $Q^{F|p}_{\tup{t}|0}$ associated with a single observable $F$ are discrete-time restrictions of a complex-valued measure $\mathcal{Q}^{F|p}\sqm{f}$ on the space of bi-trajectories, $t\mapsto (f^+(t),f^-(t))\in\Omega(F)\times\Omega(F)$,
    \begin{align*}
        Q^{F|p}_{\tup{t}|0}(\tup{f}^+,\tup{f}^-) 
        &= \iint \Big(\prod_{j=1}^n\delta_{f^+(t_j),f^+_j}\delta_{f^-(t_j),f_j^-}\Big)\mathcal{Q}^{F|p}\sqm{f}.
    \end{align*}
    In particular, due to the link~\eqref{prop:bi-prob:F-link} with probability distributions for sequential measurements, we also obtain that
    \begin{align*}
        P_{\tup{t}|0}^{F|p}(\tup{f}) &= \iint \Big(\prod_{j=1}^n \delta_{f^+(t_j),f_j}\delta_{f^-(t_j),f_j}\Big)\mathcal{Q}^{F|p}\sqm{f},
    \end{align*}
    showing that the observed probabilities $P^{F|p}_{\tup{t}|0}$ can be interpreted as the result of a superposition of \textit{interfering} pairs of elementary trajectories---the \textit{bi-trajectory picture}.
\end{enumerate}

Naturally, the bi-trajectory measure $\mathcal{Q}^{F|p}$ is not a phenomenological quantity, as it cannot be directly or indirectly measured by a classical observer. Rather, it is a logically inferred \textit{master object} of the formalism that, on the one hand, describes the dynamics of observable $F$ in terms of bi-trajectory, and on the other hand, is a ``source'' of all observable probabilities regarding $F$. However, as we demonstrate in the second part of the paper, $\mathcal{Q}^{F|p}$ turns out not to be the true master object: further deduction will lead us to the conclusion that, in this hierarchy of objects, the bi-trajectory measure associated with a single observable $F$ is not on the very top.

\section{Conclusions}
In the first part of the paper, we analyzed the outcomes of various experiments that can be reduced to the sequential deployment of devices measuring quantum observables. These outcomes are quantitatively described by phenomenological multi-time probability distributions, estimated using data obtained from many repetitions of each experiment. An immediate conclusion is that the theory describing the analyzed experimental outcomes must be \textit{non-classical} because we found that the measured sequences cannot be considered as sampling of a \textit{trajectory} representing system's observables. The breakdown of this trajectory picture results from the violation of the consistency condition by the phenomenological multi-time probabilities---a feature fundamentally incompatible with any kind of classical theory. The non-classical nature of the investigated systems manifests in various observed phenomena, including quantum interference, quantum Zeno effect, and uncertainty relations between the measured observables. Each of these phenomena corresponds to specific properties of the multi-time probabilities, providing clues for our subsequent deduction of the formalism.

We began the deduction by acknowledging that, in the case of quantum theory, it is impractical\---if not impossible\---to construct a formalism that exclusively utilizes phenomenological quantities. Therefore, to advance the program, we based the emerging formalism on the complex-valued bi-probability distributions on the space of sequence pairs. These distributions, initially introduced as a parameterization of the pair-wise interference decomposition of the phenomenological probability distributions, are now given a more fundamental role. Although bi-probability distributions are ostensibly an \textit{ad hoc} inclusion, we argue that their addition is justified. This particular formal extension is both \textit{minimal} and necessary, because its function cannot be reasonably fulfilled by the available phenomenological elements. 

Consequently, the formalism takes shape as one that describes quantum system in terms of a complex-valued measure on the space of trajectory \textit{pairs}. Thus, we arrive at an elegant and evocative explanation why quantum theory does not adhere to the classical trajectory picture: the \textit{uni}-trajectory picture is invalid because, in reality, quantum mechanics is a \textit{bi}-trajectory theory.

\section*{Acknowledgments}

D.L. acknowledges financial support by Friedrich\-/Alexander\-/Universit\"at Erlangen\-/N\"urnberg th\-rough the funding program ``Emerging Talent Initiative'' (ETI), and was partially supported by the project TEC-2024/COM-84 QUITEMAD-CM. D.C. was supported by the Polish National Science
Centre project No. 2024/55/B/ST2/01781.

\bibliographystyle{quantum}
\bibliography{bib_phenomenon}

\end{document}